\documentclass[entropy,article,accept,moreauthors,pdftex]{Definitions/mdpi} 
\usepackage{graphicx}
\usepackage{amsmath}
\usepackage{amsfonts}
\usepackage{amssymb}
\usepackage{mathtools}
\firstpage{1} 
\pubvolume{xx}
\issuenum{1}
\articlenumber{5}
\pubyear{2020}
\copyrightyear{2020}

\history{Received: date; Accepted: date; Published: date}

\Title{Complexity in economic and social systems: cryptocurrency market at around COVID-19}

\Author{Stanisław Drożdż $^{1,2}$\orcidA{}, Jarosław Kwapień $^{1}$\orcidB{}, Paweł Oświęcimka $^{1,3}$\orcidC{}, Tomasz Stanisz $^{1}$\orcidD{} and Marcin Wątorek $^{2}$\orcidE{}}

\AuthorNames{Stanisław Drożdż, Jarosław Kwapień, Paweł Oświęcimka, Tomasz Stanisz and Marcin Wątorek}

\address{%
$^{1}$ \quad Complex Systems Theory Department, Institute of Nuclear Physics, Polish Academy of Sciences, ul.~Radzikowskiego 152, 31-342 Krak\'ow, Poland\\
$^{2}$ \quad Faculty of Computer Science and Telecommunication, Cracow University of Technology, ul.~Warszawska 24, 31-155 Krak\'ow, Poland\\
$^{3}$ \quad Faculty of Physics, Astronomy and Applied Computer Science, Jagiellonian University, ul. prof. Stanisława Łojasiewicza 11, 30-348 Kraków, Poland}

\corres{Correspondence: stanislaw.drozdz@ifj.edu.pl}

\abstract{Social systems are characterized by an enormous network of connections and factors that can influence the structure and dynamics of these systems. Among them the whole economical sphere of human activity seems to be the most interrelated and complex. All financial markets, including the youngest one, the~cryptocurrency market, belong to this sphere. The~complexity of the cryptocurrency market can be studied from different perspectives. First, the~dynamics of the cryptocurrency exchange rates to other cryptocurrencies and fiat currencies can be studied and quantified by means of multifractal formalism. Second, coupling and decoupling of the cryptocurrencies and the conventional assets can be investigated with the advanced cross-correlation analyses based on fractal analysis. Third, an internal structure of the cryptocurrency market can also be a subject of analysis that exploits, for example, a network representation of the market. In~this work, we~approach the subject from all three perspectives based on data from a recent time interval between January 2019 and June 2020. This period includes the peculiar time of the Covid-19 pandemic; therefore, we~pay particular attention to this event and investigate how strong its impact on the structure and dynamics of the market was. Besides, the~studied data covers a few other significant events like double bull and bear phases in 2019. We show that, throughout the considered interval, the~exchange rate returns were multifractal with intermittent signatures of bifractality that can be associated with the most volatile periods of the market dynamics like a bull market onset in April 2019 and the Covid-19 outburst in March 2020. The~topology of a minimal spanning tree representation of the market also used to alter during these events from a distributed type without any dominant node to a highly centralized type with a dominating hub of USDT. However, the~MST topology during the pandemic differs in some details from other volatile periods.}


\begin{document}

\section{Introduction}

Whether complexity of a system is viewed in the purely intuitive sense of a nontrivial order that emerges spontaneously from an overall disorder or it is grasped more formally using one of several dozen mathematical, physical, and information-theoretic measures, we~are surrounded by its signatures and face its manifestations almost everywhere. We are complex ourselves: We live in a society that is complex and we interact with others in a complex way. There is no exaggeration in a statement that our society is the most complex structure known to us in the universe. Social phenomena like the emergence of communication and cooperation, build-up of hierarchies and organizations, opinion formation, the~emergence of political systems, and the structure and dynamics of financial markets are all among the iconic examples of the real-world complexity~\cite{kwapien2012,jakimowicz2020,klamut2020}.

Specialists from such disciplines like mathematics, physics, information theory, and data science working together with econometrists, sociologists, quantitative linguists, and psychologists for more than a quarter century have already been dealing with such phenomena trying to describe them in a language of exact science, and to model and explain them using methods and tools that had earlier been applied successfully to natural systems. while~much has already been done and much has been achieved, the~complexity of the social and economic systems is still far from being properly understood. This is why every possible effort and every meaningful contribution is welcome as it can bring us closer to the ultimate goal of understanding complexity both in reference to these systems in particular and as a physical phenomenon in general. It is also important to approach the problem from different angles by collecting many interdisciplinary works and views in one place like this Special Issue as human society eludes any narrow-scope, single-discipline analysis.

\subsection{Money, Fiat Currencies, and Cryptocurrencies}

Among a variety of emergent phenomena that we observe in human society, one of the most important is money. It appeared spontaneously and independently in many cultures and, although it used to have different material forms in different regions, it always served the same purpose: To~facilitate trade by avoiding a problem of double coincidence of needs that restricts barter trading severely and inherently. According to economical models, a status of money is acquired in a process of the spontaneous symmetry breaking by a commodity that is the most easily marketable or, in other words, that is the most liquid one~\cite{bak2001,oswiecimka2015}. After receiving such a status by some commodity, its liquidity is amplified by a kind of self-propelling mechanism, because everybody desires to have an asset that is considered as the most desirable by others. However, there is another condition for a commodity to be used as money: Its value expressed in other assets has to be viewed as stable. Sometimes it happens that current money loses its value which causes people to withdraw themselves from using it and to replace it with some other, more stable asset. Thus, for a given asset its status of money may either be durable or temporary. This is an important issue in contemporary economy based on fiat money that does not have any intrinsic value unlike the assets that used to play a role of money earlier in history. Value of the fiat currencies depends crucially on policies of the central banks, which~can be subject to change. Moreover, the~central banks may increase money supply at any time, which~can lead to inflation rate increase. This undermines confidence in the official currencies and became the ignition to introduce cryptocurrencies over a decade ago.

The first cryptocurrency was proposed in 2008---Bitcoin (BTC)~\cite{Nakamoto2008}. The~idea behind it was to decouple a currency from any institution or government, while~preserving its status of a universal means of exchange, and to base a trust in this currency solely on a technology that supports it. Such a currency had to combine the advantages of both cash and electronic money: Anonymity of use (like cash) and capability of being transferred immediately to any place in the world \mbox{(like electronic money).} The already-existing technologies of asymmetric cryptography and distributed database (with a new consensus mechanism---``proof of work'') were linked into a decentralized secure register---blockchain~\cite{Wattenhoffer} that forms a staple of BTC. Unlike traditional currencies, Bitcoin has inherently limited supply to prevent any loss of its value due to inflation.

The first widely recognized exchange enabling bitcoin to be exchanged for traditional currencies, Mt.~Gox, was launched in July 2010 followed by the first online (black) market---Silk Road. The~latter was a place where one could anonymously buy anything and pay with bitcoins, which~was the first practical application of a cryptocurrency. It significantly increased the demand and contributed to the first speculative bubble on BTC~\cite{Gerlach2018}. A subsequent crash occurred after closing Silk Road and suspending trade on Mt.~Gox between October 2013 and February 2014. As Bitcoin's recognition increased, the~use of blockchain technology became more popular and it turned out that it can also be used for trustful processing of computer codes in a decentralized way. In~2015 the Ethereum distributed computing network was launched~\cite{ethereum}, which~allows one to issue private tokens through a so-called Initial Coin Offer (ICO) and to raise capital in a simplified way for various projects. An~ICO boom that contributed to next speculative bubble on cryptocurrencies that occurred in 2017 \mbox{(the ICO-mania}~\cite{Aste2019}). At that time the number of issued cryptocurrencies doubled from 700 to 1400 and the market capitalization reached 800 billion USD. A crash in January 2018, in which BTC lost over 80\% of its value and other cryptocurrencies lost even 99\%, may be compared with the dot-com bubble crash in 2000 that ended the most euphoric phase of investor attitude towards the Internet-related companies. At present the market is more consolidated and shows signatures of maturity~\cite{PhysRep.2020}.

\subsection{Basic Information on the Blockchain Technology}

In order to create an electronic ``currency'' that can easily be exchanged for goods and operated without any central authority, while~at the same time that cannot be multiplied indefinitely like electronic files, it is required that all transactions involving that ``currency'' have to be registered publicly, which~ensures that no registry can be modified afterwards. The~Bitcoin network register consists of a sequence of block files built one upon another (a blockchain) containing information about past transactions and the instances of new Bitcoin unit creation. A new network participant has to enter the network directly via a network client or via an external wallet and must send information about the client's address and a specified Bitcoin sum it owns. This information is then distributed to all other network nodes but, in return, the~new participant is granted access to the complete information about other network node addresses and how many BTC units belong to these addresses. Thus, credibility of the system is provided by the technology itself by imposing certain set of rules each network participant must obey and by allowing the network participants to control each other. However, \mbox{since the Bitcoin} blockchain is public, one can trace the transaction history of each unit, which~in theory might compromise transaction anonymity.

The transaction correctness is guaranteed with the help of the asymmetric cryptography. Private keys of a sender and a receiver are used to encode and to decode a transaction (i.e.,~to send and to receive coins), while~their public keys are used as their public addresses allowing for their network identity verification. while~such a transaction is visible to any other network participant, nobody can effectively alter and re-encode it as they do not know the private keys of the involved parties. For~the network, in order to function correctly, the~key implemented feature is a consensus mechanism that ensures that all participants agree upon ownership of the cryptocurrency units and how many units total circulate. while~collecting information from many transactions taking place on the Bitcoin network, the~consensus mechanism has to overcome a problem that some information sources can be unreliable. It is done by the so-called {proof-of-work} (PoW) protocol used by miners, i.e.,~the network nodes with dedicated software that collect transactions, verify their correctness, and integrate them into blocks. This is a resource-consuming task so the miners are got to perform it by receiving new coins in exchange for sharing their resources with the network. The~new Bitcoin unit is generated only after majority of the miners agree upon correctness of the new block and it has been distributed over the network. The~block has to meet relevant criteria expressed by a specific form of the hash function to be considered as a valid one and included in the blockchain. Each miner decides to include a given block into its own blockchain copy individually and the consensus is settled in a kind of game with a Nash equilibrium state. One has to believe that majority of other miners agrees on the specific block's validity and adds it to its own blockchain or will not receive the profit otherwise.

Mining a new Bitcoin unit requires much energy to be spent so the very process demands optimization of the resources used and discourage padding the blocks with fictitious information as a rejection probability for such a block by other miners is too large. Therefore it serves as a proof of work that a participant made the effort of maintaining the network, indeed. The~employed solution that each new block contains a header of the previous one practically eliminates a problem of potential modifying the past transactions---it is not viable economically since it would require rebuilding of the entire chain. The~Bitcoin protocol was designed in such a way that new blocks are formed with constant frequency, which~is achieved by adjusting the amount of the corresponding calculations needed to the network's actual computing power. Moreover, the~reward for forming a new block is halved every 210,000 blocks in order to approach quasi-asymptotically an impassable limit of 21~million Bitcoin~units.
 
The Bitcoin protocol is not static and undergoes constant modifications. A reason for this is that the protocol in its original design has some drawbacks that can challenge its security and lower comfort of its use. Among the pivotal issues is low performance (the network can handle only 5 transactions per second on average, compared to 1700 transactions per second in the Visa network), high operating costs that equal the amount of electric energy consumed by small industrialized countries (like Ireland or Denmark~\cite{elektr}), and formidable computer facility. Moreover, one of the blockchain technology advantages---the inability of making changes---may sometimes be viewed as its disadvantage if one considers the protocol correcting since it requires cloning of the entire network and abandoning the original blockchain. Up to now a mechanism of reducing transaction size and allowing to pack more transactions in a single block (called ``segregated witness'', {SegWit}) has already been implemented and work on another mechanism---``Lightning Network''---that allows for micropayments outside the main blockchain and increasing the bandwidth, is currently underway.

However, such changes are viewed by inefficient by many who prefer building alternative networks from scratch or by using only certain features of the Bitcoin protocol, while~replacing other with better solutions. Thus, over the last decade, a multitude of different protocols were proposed and implemented, which~led to introduction of new cryptocurrencies. Most of them still exploit the PoW protocol, but~its the most popular alternative is the {proof-of-stake} (PoS)~\cite{pos}. In~this algorithm miners do not exist and the block validation process is granted to some randomly chosen network nodes. Consistently, the~block formation is not rewarded with new units but rather the validator nodes are rewarded with transaction fees. Fraud is discouraged by excluding the fraudulent participants from the network and securing that, in such a case, the~reward for forming a new block is smaller than possible loss in already owned units. The~main advantage of PoS is efficiency: Because of a lack of the complicated and long calculations, no specialized user group is needed to confirm blocks and everything can be done faster than in the case of PoW. There are various versions of the PoS protocol, like the ``delegated proof of stake'' (DPoS) based on voting system engaging trustful delegated network nodes or the ``proof-of-authority'' (PoA) based on granting reputation to the validator nodes instead of cryptocurrency units and abandoning the decentralization paradigm. Main advantage of both protocols (together with their hybrid versions) is scalability---more participants mean larger transaction capacity of the related network.

\subsection{Other Applications of the Blockchain Technology}

The first Bitcoin alternative that was introduced in 2011 and managed to survive until today was Litecoin (LTC). Basically, this is a Bitcoin's clone that differs from its parent in that it has a higher average creation frequency (4 min) and a higher prospected total number of units (84 million) as well as it uses different hash function ({script} instead of {dSHA-256} used by Bitcoin). These changes allowed LTC for much smaller resource demand than BTC and made LTC be computable on standard CPUs. The~first cryptocurrency that was not based on the Bitcoin's PoW protocol was Ripple (XRP)~\cite{ripple} introduced in~2012. It was intended to be used as a method of transferring money between banks and stock markets in real time even outside national borders. In~August 2020 XRP was the third cryptocurrency in terms of capitalization. A related cryptocurrency, Stellar (XLM), also offers transactions between financial institutions, but~unlike Ripple based on a proprietary code its code is open source. Both XRP and XLM do not have a fixed supply limit and, thus, they are subject to inflation.

A separate group of cryptocurrency protocols was designed to ensure user anonymity. The~corresponding cryptocurrencies are called ``private coins'': Dash (DASH), Monero (XMR), Zcash~(ZEC), and many others. Dash uses a two-layer network with PoW and miners in the first layer and PoS and ``masternodes'' in the second one. Monero, being considered as the most secure private coin and often used by the criminal world~\cite{okupm}, provides anonymity thanks to a Ring Confidential Transactions (RingCT) where the public keys (addresses) are hidden in the blockchain~\cite{ringct}. Zcash~is based on a solution that allows one to confirm information without having to disclose it. Zcash~allows for perfect anonymity of both the sender and the recipient as well as transaction size. Since the anonymous addresses are compatible with the public ones, transactions can be made between public and hidden wallets and vice versa. DASH and ZEC have a maximum supply set in advance, while~XMR does not.

Apart from the cryptocurrencies, another important category of blockchain applications is {cryptocommodities} (together with the former called {cryptoassets}). They are automatically executed computer codes that perform certain actions if certain conditions are met. Cryptocommodities enable payments for using a decentralized computing network. The~first such cryptoasset was Ethereum---\mbox{an open-source} computing platform designed for programming decentralized applications and smart contracts that was launched in 2015~\cite{ethereum}. This platform has its own programming language and its own cryptocurrency, Ethereum (ETH), that serves as a payment unit for carrying out computational operations on the platform. Ethereum is based on PoW consensus mechanism, but~it uses another hash function ({Ethash}) supporting use of GPUs in the mining process and there is no upper limit on mining. Instead of fixed block size, here each block requires a specific number of ``Gas'' units related to the computing power needed to complete the transactions it contains. The~average block-completion frequency is 15~s and the maximum transaction number per second is around 25.~The~Ethereum concept gained quickly high popularity among the cryptocurrency community and, currently, ETH is the second crytocurrency in terms of capitalization. The~success of smart contracts (i.e.,~computer codes allowing for automatic execution and control of transaction agreement actions) and possibility of collecting funds under Initial Coin Offers on the Ethereum platform, gave a boost to the emergence of similar platforms offering possibility of creating applications in a decentralized environment. Major projects of this type include EOS and Cardano; both have their own cryptocurrencies and both allow for collecting funds under ICOs.

Yet another group of cryptoassets are {tokens}, which~are means of payment in decentralized applications built on platforms like Ethereum or contracts that are issued within ICOs for development of blockchain ventures. They usually don't have their own blockchain. In~general, the~blockchain technology, thanks to elimination of the need to trust individual participants of a given system and ensuring security, can satisfactorily be used wherever there is a central intermediary connecting sellers and buyers who earns on commissions (for example, Uber and Airbnb). Some of the already introduced applications in the token form are Augur (a platform enabling creation and participation in plants from any thematic range), Filecoin (a decentralized file storage system based on the PoW system that rewards users for sharing their computer storage devices), and IOTA (a project of a partially decentralized, open settlement platform for the needs of the so-called ``Internet of things''), Basic Attention Token \mbox{(a project} designed to connect advertisers and content creators with users that rewards the creators for attracting users with the content they provide). Finally, the~so-called \mbox{``stable coins''---a combination} of the token and cryptocurrency assets---allow one to relate their value to some other, more conventional asset like US dollar (e.g., USDT, USDC, TUSD, or PAX).

\subsection{Cryptocurrency Market}

Cryptocurrency trading is possible, because they are easily convertible to traditional currencies like USD or EUR and to other cryptocurrencies. This possibility is provided by 330 trading platforms (August~2020) open 24~h a day, seven days a week. This, together with a fact that the most investors are individuals, distinguishes the cryptocurrency market from Forex, where~trading takes place from Monday to Friday essentially on the OTC market where mainly banks and other financial institutions participate in. Another peculiarity of the cryptocurrency market is that there is no reference exchange rate unlike Forex, where~such reference rates are provided by Reuters. The~sole exception is Bitcoin, whose exchange rate to USD is given by futures quoted on Chicago Mercantile Exchange~\cite{CME}. Decentralization of the market means that the same cryptocurrency pairs are traded on different platforms, which---if accompanied by limited liquidity---can lead to sizeable valuation differences between platforms that produce arbitrage opportunities, both the dual and triangluar ones~\cite{arbitrage,gebarowski2019,PhysRep.2020}. 

The entire cryptocurrency market capitalization is around 350 billion USD, which~is close to the capitalization of a middle-size stock exchange and also comparable with the capitalization of the largest American companies. There are 6500 different cryptocurrencies on the market right now, which~gives a total of nearly 26,500 cryptocurrency pairs~\cite{coinmarket}. Founded in 2017, Binance~\cite{Binance} is currently one of the largest cryptocurrency exchange in terms of volume. Binance offers trading on approximately 650~cryptocurrency pairs including pairs with its own cryptocurrency called binance (BNB), used to pay commissions on this exchange.

The spectacular development of a cryptocurrency market has attracted much interest of the scientific community. The~first Bitcoin-related papers were published already in 2013--2015~\cite{Kristoufek2013,Kristoufek2015}, but~a real boom on cryptocurrency-related publications occurred after 2017. Initially, only bitcoin was of significant interest~\cite{Bariviera2017,DrozdzBTC2018,Garnier2019}, but~soon also other cryptocurrencies went under investigation~\cite{Wu2018,futinternet2019,Kristoufek2019}. Then there appeared studies reporting on correlations within the market~\cite{stosic2018,zunino2018,bouri2019,zieba2019,chaos2020,FERREIRA2020,PAPADIMITRIOU2020,polovnikov2020,medina2020}, and its relationship with regular markets~\cite{Corbet2018,corelli2018,ji2018,kristj2019,ausloos2020}. Recently, some researchers focused their attention on possible use of BTC as a hedging instrument for Forex~\cite{urquhart2019}, for gold and other commodities~\cite{shahzad2019}, as well as for the stock markets~\cite{shahzad2019a,wang2019}. There is also a few review papers devoted to the cryptocurrency markets:\cite{Corbet2019,fang2020cryptocurrency,PhysRep.2020}.

The cryptocurrency market has already gone through a long route from a mere curiosity and a playground for the technology enthusiasts, via an emerging-market stage characterized by a relatively small capitalization, poor liquidity, large price fluctuations, short-term memory, frequent arbitrage opportunities, and weak complexity, to a more mature form characterized by medium capitalization, improved liquidity, inverse-cubic power-law fluctuations~\cite{gopi1998,gabaix2003}, long-term memory, sparse arbitrage opportunities, and increasing complexity. This is the most interesting aspect of the cryptocurrency market route to maturity: The signatures of complexity that are best quantified in terms of the multifractal analysis. See Ref.~\cite{PhysRep.2020} for a comprehensive study of this transition started in 2012 and ended essentially in 2018, as viewed from the multifractality perspective. Here~we shall consider a more recent period of 2019--2020, which~comprises, among others, two significant events, i.e.,~the bull market between April and July 2019 and the Covid-19 pandemics (from March 2020). Based on high-frequency data covering a large number of cryptocurrency pairs and a few principal traditional-market assets, we~investigate a potential impact of these events on the cryptocurrency market structure and its relation to the traditional markets.


\section{Methods and Results}
\unskip
\subsection{Data}
\label{sect::data}

For this study we collected high-frequency recordings of X/BTC and BTC/USDT exchange rates, where~X is one of 128 cryptocurrencies traded on Binance platform~\cite{Binance} and USDT is related to USD by a 1:1 peg~\cite{tether}. The~exchange rates $P(t)$ were sampled every 1~min. We calculated their normalized logarithmic returns $r_{\Delta t}$ defined by
\begin{equation}
r_{\Delta t}=(R_{\Delta t}-\mu_R)/ \sigma_R, \quad R_{\Delta t}(t)=\log(P(t+\Delta t))-\log(P(t)),
\end{equation}
where $\mu_R$ and $\sigma_R$ are mean and standard deviation of $R_{\Delta t}(t)$, respectively, and $\Delta t$ is sampling interval. We also collected 1-min quotes of several conventional assets expressed in US dollar---13 currencies: AUD, EUR, GBP, NZD, CAD, CHF, CNH, JPY, MXN, NOK, PLN, TRY, ZAR, three stock market indices: Dow Jones Industrial Average (DJI), Nasdaq100, S\&P500, and four commodities: XAU (gold), CL~(crude oil), XAG (silver), and  HG (copper). They all come from Dukascopy platform~\cite{Dukascopy}, so~do the BTC/USD and ETH/USD exchange rates. These quotes were also transformed into time series~of~returns.

\subsection{Multifractal Formalism}
\label{sect::formalism}

Multifractal analysis is one of the most promising methods of studying empirical data representing natural and social systems as it is able to quantify complexity of such systems and express it in a relatively simple way with a small set of associated quantities. It has already been applied in many works to univariate and multivariate data sets from a number of different systems: Physics~\cite{subramaniam2008}, biology~\cite{ivanov1999}, chemistry~\cite{stanley1988}, geophysics~\cite{witt2013}, hydrology~\cite{koscielny2006}, atmospheric physics~\cite{kantelhardt2006}, quantitative linguistics~\cite{drozdz2016}, behavioral sciences~\cite{ihlen2013}, cognitive structures~\cite{dixon2012}, music~\cite{jafari2007}, songbird rhythms~\cite{roeske2018}, physiology~\cite{nagy2017}, human behaviour~\cite{domanski2017}, social psychology~\cite{Krawczyk2019} and even ecological sciences~\cite{stephen2013}, but~especially financial markets~\cite{ausloos2002,kwapien2005,oswiecimka2005,drozdzepps,Rak2015,grech2016,Zhao2017,Watorek2019,multirev}.

Let us consider two time series of the same length: $x_i$, $y_i$, where~$i=1,...,T$ ($T$ has to be large enough to overcome statistical uncertainties). Signal profiles are created from these time series by integrating and subtracting their mean:
\begin{equation}
X(j) =\sum_{i=1}^j[x_{i}-\langle x\rangle] ,\quad
Y(j) =\sum_{i=1}^j[y_{i}-\langle y\rangle].
\end{equation}

These signal profiles are then divided into segments $\nu$ of length $s$. They may be separate or partially overlapping; if they are separate, their number is $M_s=\lfloor T/s \rfloor $. A local trend is then removed from each segment by fitting the data with polynomials $P^{(m)}_{X,\nu}$, $P^{(m)}_{Y,\nu}$ of degree $m$ (typically, it is $m=2$~\cite{kantelhardt02,oswiecimka2006,oswiecimka2013}). Now covariance $F_{xy}^2$ is determined from the residual signals for each segment~\cite{podobnik2008,zhou2009}:
\begin{equation}
F_{xy}^{2}(\nu,s) = \frac{1}{s} \sum_{k=1}^{s} \lbrace \left[ X((\nu-1)s+k)-P^{(m)}_{X,\nu}(k) \right] \left[ Y((\nu-1)s+k)-P^{(m)}_{Y,\nu}(k) \right] \rbrace
\label{Fxy2}
\end{equation}
and then it is used to calculate the $q$-th order fluctuation function~\cite{oswiecimka2014}:
\begin{equation}
F_{xy}^{q}(s)=\frac{1}{M_s}\sum_{\nu=1}^{M_s} {\rm
sign}(F_{xy}^{2}(\nu,s))|F_{xy}^{2}(\nu,s)|^{q/2},
\label{Fq}
\end{equation}
where ${\rm sign}(F_{xy}^{2}(\nu,s))$ means a sign function. If $F_{xy}^2(\nu,s)$ is considered as a value of a random variable, the~parameter $q$ resembles an exponent specifying the order of the moment: Its large positive values favour segments characterized by large variance by increasing their relative magnitude with respect to small-variance segments, while~negative values of $q$ do the opposite. Thus, by applying different values of $q$, one can construct effective filters that select the segments of a certain variance range.

The fluctuation function (\ref{Fq}) has to be calculated for different segment lengths $s$. If $F_{xy}^{q}(s)$ is of a power-law form, i.e.,~
\begin{equation}
F_{xy}^{q}(s)^{1/q}=F_{xy}(q,s) \sim s^{\lambda(q)}, 
\label{Fxy}
\end{equation}
where $q \neq 0$, the~original time series $x_i$ and $y_i$ are fractally cross-correlated. If $\lambda(q)={\rm const}$, this is monofractal cross-correlation, otherwise it is multifractal one.

A special case is $x_i \equiv y_i$ for all $i$ (one signal). In~this case we have
\begin{equation}
F(q,s)=\Big[\frac{1}{M_s}\sum^{M_s}_{\nu=1}{[F^2(\nu,s)]^{\frac{q}{2}}}\Big]^{\frac{1}{q}}
\label{F}
\end{equation}
and the fractal case corresponds to
\begin{equation}
F(q,s) \sim s^{h(q)},
\label{Hq}
\end{equation}
where $h(q)$ is the generalized Hurst exponent. For~$h(q)={\rm const}$ the signal is monofractal, otherwise it is multifractal~\cite{kantelhardt02}. A useful measure of fractal properties is singularity spectrum $f(\alpha)$ defined by
\begin{equation}
\alpha=h(q)+qh'(q), \quad f(\alpha)=q[\alpha-h(q)]+1,
\label{spektrum}
\end{equation}
where $\alpha$ is the H\"older exponent. $f(\alpha)$ can be interpreted as a fractal dimension of
the singularities characterized by a given $\alpha$. In~the monofractal case it consists of a single point, while~in the multifractal case it can have a shape of inverted parabola or some asymmetric concave function.

Width of $f(\alpha)$ can be interpreted as a measure of a signal's complexity, because the wider it is, the~more singularity types can be identified in this signal. This width depends on a range of $q$ and it is quantified by
\begin{equation}
\Delta \alpha = \alpha_{\rm max}  - \alpha_{\rm min},
\label{Dm}
\end{equation}
where $\alpha_{\rm min}=\alpha(q_{\rm max})$ and $\alpha_{\rm max}=\alpha(q_{\rm min})$ are the minimum and maximum value of $\alpha$ that have been calculated for different values of $q$. Another important feature of the $f(\alpha)$ its left-right asymmetry~\cite{drozdz2015}. A left-hand-side asymmetry corresponds to more diverse multifractality (stronger correlations) at the large amplitude level, while~a right-hand-side asymmetry indicates that signal parts with small amplitude are a dominant source of multifractality.

As $F^q_{xy}(s)$ denotes the $q$th-order detrended covariance, one can define the $q$th-order detrended correlation coefficient~\cite{zebende2011,kwapien2015}:
\begin{equation}
\rho(q,s) = {F_{xy}^q(s) \over \sqrt{ F_{xx}^q(s) F_{yy}^q(s) }},
\label{rhoq}
\end{equation}
in analogy to the $q$th-order Pearson correlation coefficient. Here~$F_{xx}$ and $F_{yy}$ are calculated from Equation~(\ref{F}). The~coefficient $\rho(q,s)$ can assume values in a range $[-1,1]$ provided $q>0$. For~$q \le 0$ a situation becomes more complicated as $\rho(q,s)$ may fall outside that range, which~requires more delicate interpretation~\cite{kwapien2015}. Therefore, many studies in which $\rho(q,s)$ is used are carried out with a restriction $q>0$. The~coefficient $\rho(q,s)$ describes detrended cross-correlations between two signals on different scales $s$ after amplifying data points within a given amplitude range. This filtering ability of $\rho(q,s)$ is its advantage over more standard correlation measures, because the cross-correlation strength among empirical time series can be size-dependent~\cite{kwapien2017}. The~coefficient $\rho(q,s)$ may be used for any two signals without a requirement that they have to be fractal.

\subsection{Multifractal Properties of the Cryptocurrency Market}
\label{sect::multifractal}

\begin{figure}[H]%
\centering
\includegraphics[width=0.8\textwidth]{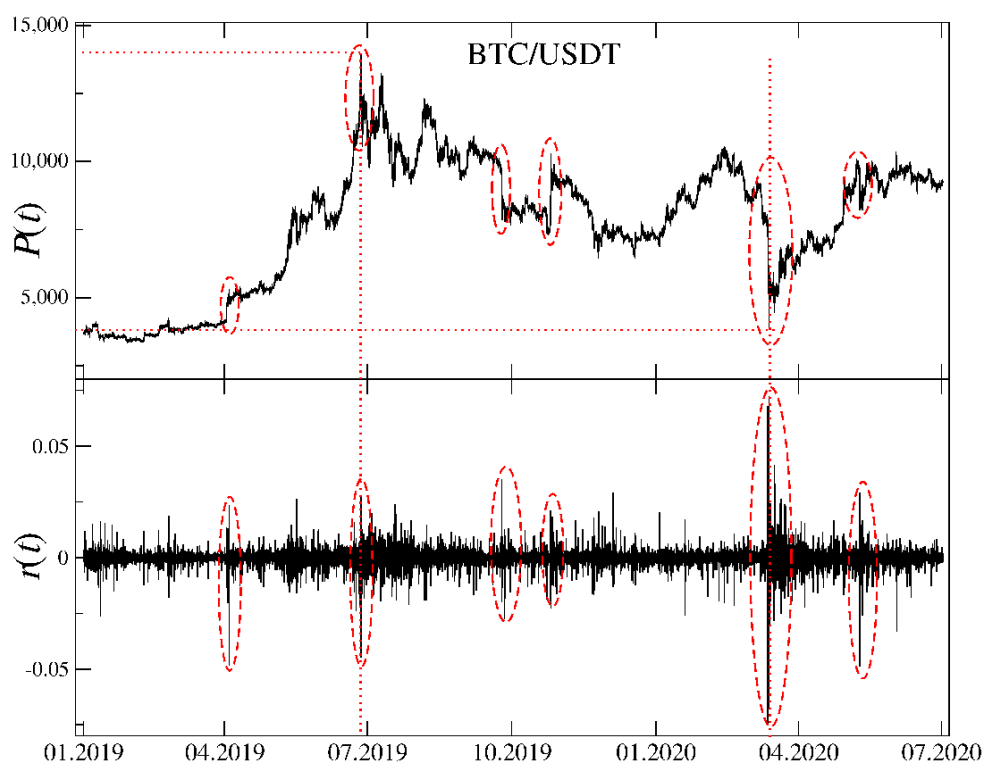}
\caption{Time evolution of the BTC/USDT exchange rate (top) together with the corresponding logarithmic returns (bottom). Several interesting events can be distinguished like start of a bull market in April 2019 and its end in July 2019, a sudden decrease and then an equally sudden increase in October and November 2019, the~Covid-19 pandemic outbreak and related panic in March 2020 and the pandemic's 2 wave in June 2020. Local extrema of $P(t)$ are indicated by the vertical (time) and horizontal (price) dotted lines.}
\label{fig::BTC}
\end{figure}

We start our analysis by taking a look at the BTC/USDT exchange rate from 01/2019 to 06/2020. This period shown in Figure~\ref{fig::BTC} (top panel) starts near the lowest point of the bear market ($\sim$3200 USDT) that begun in 12/2017 and deprived BTC over 80\% of its maximum value. During the subsequent one a half year BTC/USDT rate experienced a growth to a local maximum of 12,800 USDT in 07/2019 (+300\%), a local minimum in 03/2020 at 4400 USDT (over 60\% loss) related to a Covid-19 pandemic onset, and a recent growth to a present price of 12,200 USDT (+170\%). The~BTC price expressed in USDT was highly unstable over the considered period. This observation is supported by Figure~\ref{fig::BTC} (bottom panel) that shows BTC/USDT 1~min returns.

Despite of the fact that BTC/USDT rate is the most important observable on the cryptocurrency market since BTC has the largest capitalization, it cannot be used as a proxy allowing one to describe dynamics of the whole market, which~is in fact much richer. Thus, in order to express the evolution of a significant part of the market in terms of a single quantity, a market index was created from the exchange rates X/USDT (with X standing for a cryptocurrency) for 8 the most capitalized cryptocurrencies: BTC, ETH, XRP, BCH, LTC, ADA, BNB, and EOS. In~2020, these assets stand for 88\% of the market capitalization. In~order to create the index, the~exchange rates were summed with the same weight despite the difference in capitalization.~A parallel, weighted index would predominantly reflect the dynamics of BTC, ETH, and XRP, so we prefer the unweighted version as more a diversified~one.

\begin{figure}[H]%
\centering
\includegraphics[width=0.8\textwidth]{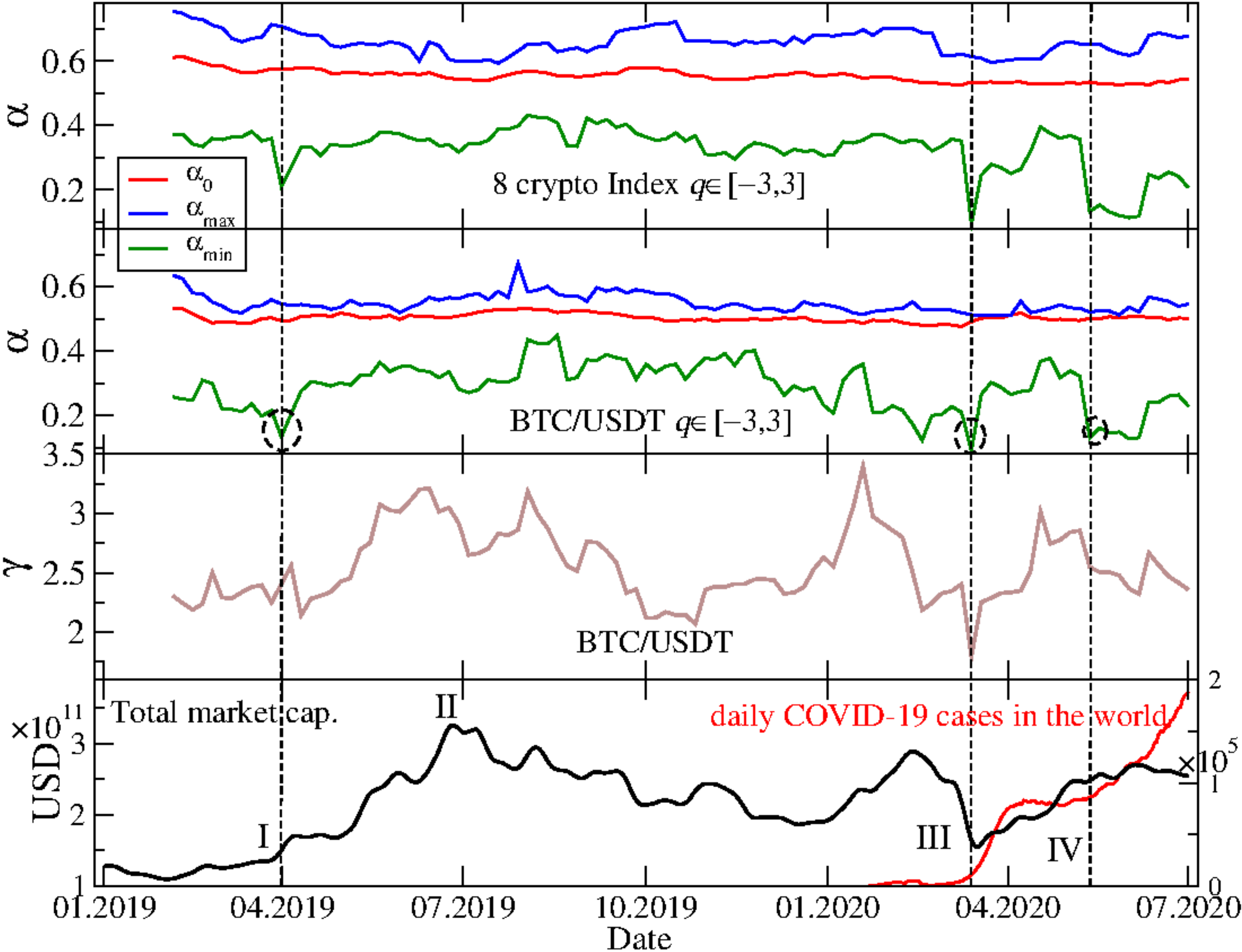}
\caption{(Top) Characteristic values of the H\"older exponent: $\alpha_{\rm min}$ (green line, bottom), $\alpha_0$ (red line, middle), and $\alpha_{\rm max}$ (blue line, top)---see Equation~(\ref{Dm}) in Section~\ref{sect::formalism} and Figure~\ref{fig::falpha}---describing the singularity spectra $f(\alpha)$ for the index returns representing 8 the most capitalized cryptocurrencies, calculated in a 30-day-long moving window with a step of five days and for $-3 \le q \le 3$. Each date represent a window that ends on that day. (Upper middle) The same quantities as in the top panel, but~here calculated for the BTC/USDT exchange rate returns. Three interesting cases of small $\alpha_{\rm min}$ are indicated by dashed circles. (Lower middle) Scaling exponent $\gamma$ of the cumulative distribution function fitted to tails of the empirical cdf in each moving window position. Values equal or below $\gamma=2$ correspond to L\'evy-stable distributions. (Bottom) Total cryptocurrency market capitalization and new Covid-19 cases in the world as function of time. Characteristic events are indicated by vertical dashed lines and Roman numerals: Start of a bull market in April 2019 (event I), its end in July 2019 (event II), the~Covid-19 panic in March 2020 (event III), and start of the 2nd wave of the pandemic in May-June 2020 (event IV).}
\label{fig::MFDFA.index}
\end{figure}

\begin{figure}[H]%
\centering
\includegraphics[width=0.7\textwidth]{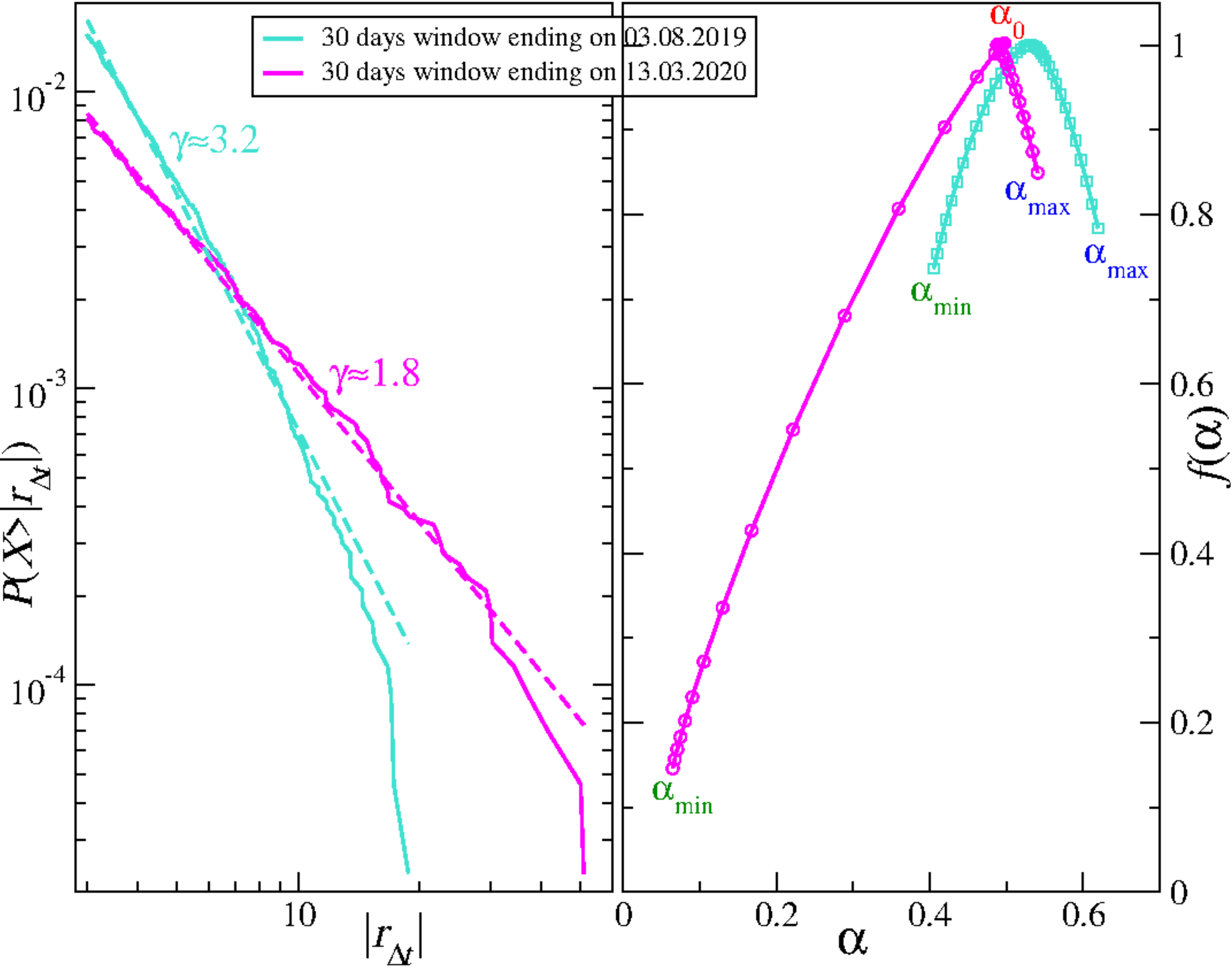}
\caption{(Left) Cumulative distribution function $P(X>|r_{\Delta t}|)$ calculated in 30-day windows. Two~extreme cases of power-law tail are shown with the scaling exponent $\gamma \approx 1.8$ (mid February--mid March 2020) and $\gamma \approx 3.2$ (July 2019) representing stable and unstable distributions, respectively. (Right) Singularity spectra $f(\alpha)$ calculated in the same windows as above. An~example of asymmetric, bifractal-like spectrum (mid February - mid March 2020) and an example of symmetric spectrum \mbox{(July 2019)} are shown together with characteristic values of the H\"older exponent: $\alpha_{\rm min}$, $\alpha_0$, and $\alpha_{\rm max}$ (see Equation~(\ref{Dm}) in Section~\ref{sect::formalism}).}
\label{fig::falpha}
\end{figure}

Figure~\ref{fig::MFDFA.index} shows results of the multifractal analysis of the cryptocurrency index returns and the BTC/USDT returns performed by using a moving window of 30 days with a 5-day step. Instead~of presenting the singularity spectra $f(\alpha)$ for each window position, temporal evolution of the key quantities describing shape of these spectra is shown: $\alpha_{\rm min}(t)$, $\alpha_0(t)$, and $\alpha_{\rm max}(t)$ (see right panel of Figure~\ref{fig::falpha} for the examples). These quantities allow for inferring about the singularity spectrum localization, width, and possible asymmetry of its shoulders~\cite{drozdz2018}. We restricted the applied values of $q$~to $[-3,3]$ for a reason that will be explained later. By looking at the spectra for BTC/USDT \mbox{(the second topmost} panel in Figure~\ref{fig::MFDFA.index}), one sees that a difference $\Delta \alpha = \alpha_{\rm max} - \alpha_{\min}$ describing the spectrum width is sufficient to infer about multifractality of the data under study. This agrees with results of our previous study~\cite{PhysRep.2020}.

Except for July-August 2019, when $f(\alpha)$ is left-right symmetric ($\alpha_{\rm max} - \alpha_0 \approx \alpha_0 - \alpha_{\rm min}$), throughout the remaining part of the analyzed period there is significant asymmetry with the left-hand shoulder ($q > 0$) being much longer than the right-hand one ($q<0$). In~a few instances, i.e.,~in April 2019, January 2020, March 2020, and May-June 2020, this asymmetry of $f(\alpha)$ became extreme and revealed a bifractal-like shape (see also~\cite{mnif2020}). Mathematical bifractals are characterized by the existence of only 2 singularity types with $\alpha_1=0$ and $0 < \alpha_2 < 1$. However, in practical situations, the~finite-size effects smear the spectra so that in such a case there is a continuous transition between both singularity types and a spectrum consists of a long left shoulder reaching a vicinity of $\alpha_1=0$ and a residual right shoulder near $\alpha_2$~\cite{EPL2010,drozdz2015}. Two~characteristic cases of $f(\alpha)$ (symmetry and bifractal-like asymmetry) are~shown in Figure~\ref{fig::falpha} (right panel).

On the probability distribution function level, the~actual bifractal spectra occur if a signal under study has a heavy-tailed pdf in the L\'evy-stable regime ($p(|r_{\Delta t}|) \sim 1 / |r_{\Delta t}|^{\gamma+1}$, where~$\gamma \le 2$), but~empirically one can sometimes obtain a strong left-hand-side asymmetry even for a signal with an unstable pdf provided it is substantially leptokurtic~\cite{EPL2010}. Figure~\ref{fig::falpha} illustrates this connection between a cumulative distribution function (cdf) $P(X>|r_{\Delta t}|) \sim 1/|r_{\Delta t}|^{\gamma}$ (left panel) and $f(\alpha)$ (right panel) for two time windows that show clearly different properties of both cdf and $f(\alpha)$---a symmetric $f(\alpha)$ corresponding to a steep cdf with $\gamma \approx 3.2$ (a window covering July and August 2019) and an asymmetric $f(\alpha)$ corresponding to a heavy-tail cdf with $\gamma \approx 1.8$ (a window covering February and March 2020). These values of $\gamma$ point to the aforementioned restriction $-3 \le q \le 3$ applied to the calculation of $F_{xy}^q$ and $\Delta\alpha$: For $|q|>3$ the moments of the distribution $p(|r_{\Delta t}|)$ can diverge, so can $F_{xy}^q(s)$ especially for small scales $s$.

The BTC/USDT return distribution function reflects a combination of two factors:~(1)~How fast the information spreads over the market---the heavier tails are, the~slower this spreading proceeds, and (2)~how volatile is the market---periods that cover turmoils with high volatility also result in heavier tails of pdf/cdf. It was documented in Ref.~\cite{PhysRep.2020} that along with a process of the cryptocurrency market maturation the scaling exponent $\gamma$ increases with time. This happens because as recognition of the market and its capitalization increase, more and more transactions take place, which~decreases the average inter-transaction waiting time and allows the market participants to react faster. Faster~reactions are crucial for the market to become efficient, which~means more Gaussian-like fluctuations (larger $\gamma$). On the other hand, extremely large fluctuations and amplified volatility are characteristic for the periods with negative events, which~decrease $\gamma$. Figure~\ref{fig::MFDFA.index} (the 3rd panel from top) shows a scaling exponent $\gamma$ obtained by fitting a power-law function to the BTC/USDT returns cdf in each position of the 30-day moving window. Indeed, such events like a bear market after July 2019 and the Covid-19 outbreak in March 2020 resulted in relatively small values of $\gamma$, while~a bull market between April and July 2019 and an escape from conventional assets to alternative ones observed between January and February 2020 led to larger values of $\gamma$.

Even if BTC is only one of many actively traded cryptocurrencies on the Binance platform, its~strongest position due to the largest capitalization (between 50\% and 70\% of total market capitalization in the considered period) causes other cryptocurrencies to evolve accordingly. This`observation comes from the topmost panel of Figure~\ref{fig::MFDFA.index} presenting $\alpha_{\rm min}$, $\alpha_0$, and $\alpha_{\rm max}$ for the 8-cryptocurrency index. Qualitatively, the~temporal course of these quantities does not differ much from the temporal course of their counterparts for BTC/USDT (the 2nd panel from top). The~only significant difference is that for the index a transition to a bifractal-like $f(\alpha)$ spectrum in March 2020 was sharp and it was not preceded by its gradual change starting from January 2020 as it was the case with BTC/USDT.

By looking at the bottom panel of Figure~\ref{fig::MFDFA.index}, where~total market capitalization is plotted as a function of time together with the Covid-19 pandemic severity parametrized by the number of daily new cases, and by comparing this plot with the remaining three, one can infer about how various market events and the pandemic influenced complexity of the market dynamics. The~main events are denoted by Roman numerals: The beginning of the bull market in April 2019 (event I), its end in July 2019 (event II), the~Covid-19 panic in March 2020 (event III), and the second pandemic wave that started in May 2020 (event IV). These events could be distinguished because they were associated with particularly large fluctuations (Figure~\ref{fig::BTC}). Among them, the~events I, III, and IV had a significant impact on the multifractal properties of the exchange rate fluctuations by sizeable decreasing of $\alpha_{\rm min}$ (visible both for the cryptocurrency index and the BTC/USDT exchange rate). However, the~event II did not have such an impact. In~contrast, the~pdf/cdf tails reflected overall market phase more than specific events except for the Covid-19 panic in March 2020.

\subsection{Cryptocurrency Market Versus Standard Markets}
\label{sect::cross-market}

From a practical point of view, among the most interesting issues related to any asset and any market is how much it is related to other assets or markets, and, in other words, whether it can be exploited for portfolio diversification and hedging~\cite{demir2020,conlon2020,kristoufek2020}. As the investors may be interested in different time horizons and may want to hedge against events of different magnitude, the~$q$-dependent detrended cross-correlation coefficient $\rho(q,s)$ defined by Equation~(\ref{rhoq}) is a measure that is particularly useful in this context since it is sensitive to both scale and amplitude of the asset price returns. We~choose the BTC/USD exchange rate as a representative of the whole cryptocurrency market---it is the most frequently traded asset, the~most capitalized asset, and the most mature one (based on our previous results~\cite{PhysRep.2020}). We calculate $\rho(q,s)$ for this rate and each of the remaining conventional assets listed in Section~\ref{sect::data}. However, we~observe that this measure behaves similar for S\&P500, Nasdaq100, and DJI, so we abandon the latter two indices and show only the results for S\&P500. In~parallel, we~neglect AUD, NZD, ZAR, CHN, MXN, EUR, GBP, NOK, TRY, and PLN as their correlations with BTC were close to zero throughout the period under consideration. We consider two temporal scales that correspond to different horizons: $s=10$~min, which~is the shortest scale available provided we use 1-min returns, and $s=360$ min that represents approximately a trading day in the US stock market. The~latter value means that in a moving widow there was only 10 segments over which the averaging was carried out in $F^q_{xy}(s)$ (see Equation~(\ref{Fq})), so we could not look at longer scales. As regards the parameter $q$, we~focused on $q>0$ in order to avoid the interpretation subtleties that could occur otherwise (see Section~\ref{sect::formalism}). We carried out our analysis for different values of $q$, but~here we shall report only the results for $q=1$ and $q=4$. The~former choice did not favour any value range of the fluctuation function $F_{xy}^2$ since, for each segment $\nu$ in Equation~(\ref{Fq}), it was counted with the same weight. Therefore $q=1$ allowed us for considering all time periods in the same way irrespective of whether the market was quiet or turbulent. On the other hand, $q=4$ corresponds to favouring the segments with the largest return covariance and degrading the other segments. Thus, this case is interesting from a perspective of the investors that want to hedge against the largest price movements and the largest-impact events. The~intermediate values of $q$ were also investigated, but~the related results fell between these two cases and, thus, they are not presented here. Moreover, the~calculations for $q>4$ were progressively less interesting with increasing $q$ as the event statistics became poor.

\begin{figure}[H]%
\centering
\includegraphics[width=0.8\textwidth]{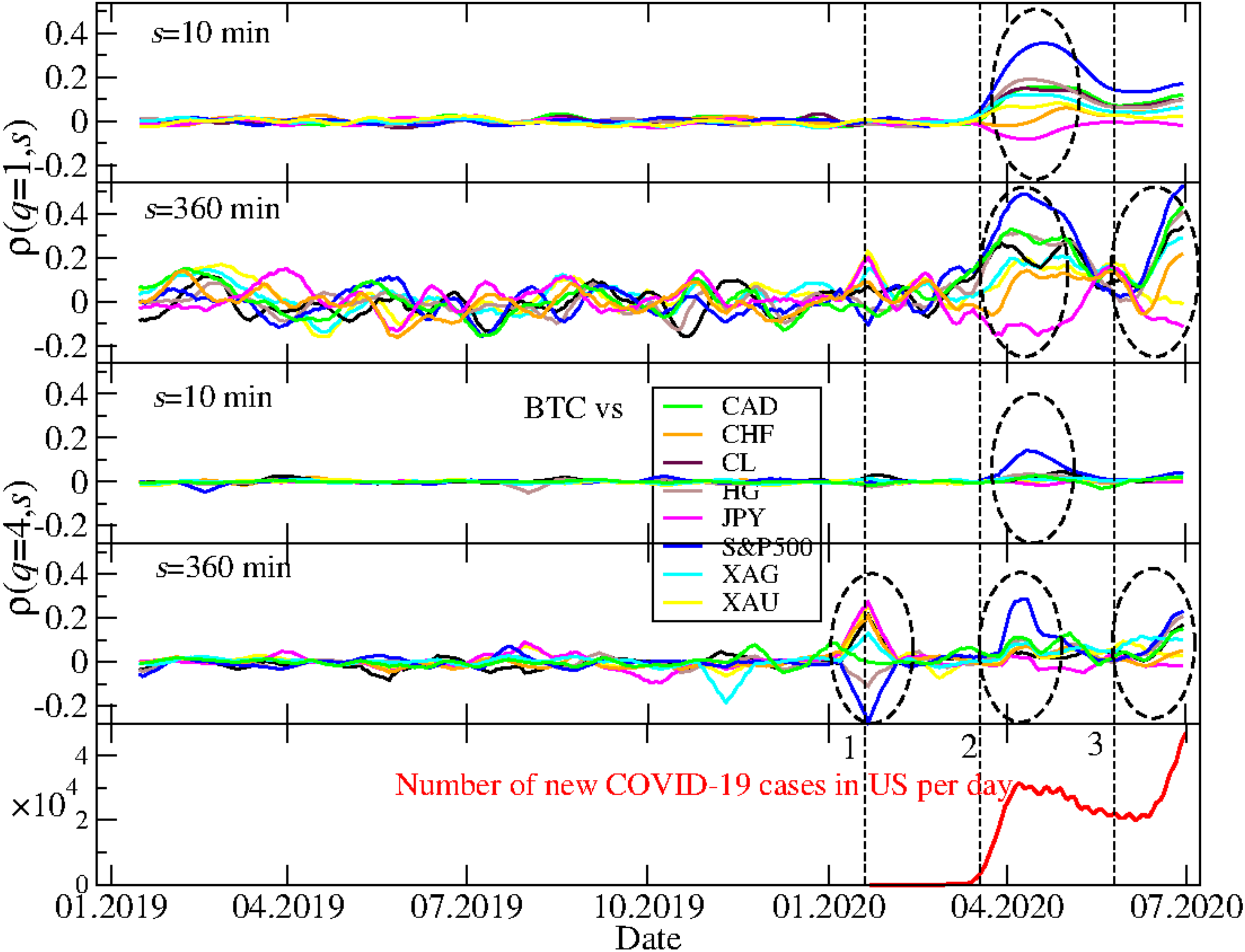}
\caption{Temporal evolution of the detrended cross-correlation coefficient $\rho(q,s)$ calculated for the BTC/USD exchange rate and the conventional assets expressed in US dollar: Japanese yen (JPY), Canadian dollar (CAD), Swiss franc (CHF), crude oil (CL), silver (XAG), gold (XAU), copper (HG), and~the S\&P500 index. The~$\rho(q,s)$ coefficient was calculated in a moving 10-day-long window with a step of 1 day and its $s$ and $q$ parameters are represented by $s=10$ min (the shortest scale), $s=360$~min (approximately a trading day in the US stock market), $q=1$ (all data points are considered), and $q=4$ (only the data points with large amplitude are considered). In~each panel events with the statistically significant, genuine cross-correlations are marked with dashed ellipses. The~daily number of new Covid-19 cases in the United States is also shown for a comparison (bottom). The~particular market events are indicated: (1) A sharp drop of the US stock market indices after the first case of Covid-19 had been identified in the United States; (2) a Covid-19 outburst related panic on the financial markets; (3) a bear market return on risky assets that was related to the 2nd wave of the pandemic.}
\label{fig::rhoq.BTC}
\end{figure}

Figure~\ref{fig::rhoq.BTC} displays temporal course of $\rho(q,s)$ for a combination of the above-described cases of \mbox{$s$ and $q$.} In each panel the cross-correlation coefficients for BTC and each of the 8 other assets are shown. Curiously, we~do not observe any statistically significant values of $\rho(q,s)$ during the whole year 2019 even though there were then important events on the cryptocurrency market, like the bull and the subsequent bear market. However, these events were not related to any of the conventional assets considered here. We see that even the periods of high volatility in April and July-August 2019 did not cause any action that could potentially be sensed by the regular markets. We can explain this lack of reaction by a relatively small capitalization of the cryptocurrency market---far too low for the other markets to detect a possible influx of a capital withdrawn from cryptocurrencies (if such an influx actually took place). In~2019 there was no turmoil in the conventional markets, thus nothing could correlate the cryptocurrencies with the conventional assets from this direction, too.

\newpage
In contrast, there were 1 to 3 periods of significant inter-market cross-correlations in the first half of 2020, dependent on $s$ and $q$. The~first important period in the end of January and the begin of February was associated with a sharp drop of S\&P500 and other US stock market indices triggered by the first identified local case of Covid-19. The~cryptocurrency market reacted rather moderately with only a short period of large and delayed fluctuations. This is why there are no significant elevation of $|\rho(q,s)|$ for short scales for any return size. The~cryptocurrency market must have been calm enough to delay reaction so long that it is identifiable only on large scales (like $s=360$ min). The~cross-correlation is positive with the fiat currencies, while~negative with the US stock markets. As all the considered assets are expressed in USD, the~positive correlations of BTC with the fiat currencies in January/February 2020 mean that there was a global flee from US dollar to other major currencies that increased the corresponding exchange rates as well as a flee from the US stock markets to the cryptocurrency market.

Opposite situation took place during the pandemic's 2nd wave in June 2020 (and, possibly, beyond that month): The cross-correlations are stronger for $q=1$ than for $q=4$. On the one hand, for $s=10$ min moderate values of $\rho(q,s)$, mainly positive ones, are seen for $q=1$, but~they are not seen for $q=4$. On the other hand, for $s=360$ min large values of $\rho(q,s)$ are observed for $q=1$ and slightly smaller, but~also significant, for $q=4$. Therefore we still see that the correlations cannot be built in their full magnitude on short scales and they need some time to develop completely. However, a larger $\rho(q,s)$ for $q=1$ indicates that the cross-correlations affect all the returns irrespective of their amplitude (small and moderate returns dominate in number, thus more segments in $F^q_{xy}(s)$ are correlated in this case than in the case of $q=4$). Majority of assets are correlated positively with an exception for JPY that is anticorrelated with BTC.

The third and the most important interval of the inter-market cross-correlations happened between the two above discussed events and it covers the pandemic outbreak and a financial market panic in March 2020. The~mutual coupling of the different markets (including the BTC/USD exchange rate for the first time) was especially evident in this case. We observe the same rule here as in the two previous cases that the cross-correlations need time to build up, thus they are stronger for $s=360$ min than for $s=10$ min. However, they are clearly evident even for $s=10$ min. Interestingly, if we look at the largest returns ($q=4$), we~see that BTC is (positively) correlated mainly with S\&P500, while~it is more independent as regards other assets. If we take a look at the results for $q=1$, the~cross-correlations appear strong between BTC and all other assets except for CHF (only small negative correlation) and gold (XAU). The~corresponding values of $\rho(q,s)$ are positive for S\&P500, CAD, copper (HG), crude oil (CL), and silver (XAG), while~they are negative for JPY. This cannot be viewed as a surprise since the Swiss franc and Japanese yen are considered safe assets together with gold and their pricing in USD behave differently than the remaining assets' pricing did.

\begin{figure}[H]%
\centering
\includegraphics[width=0.8\textwidth]{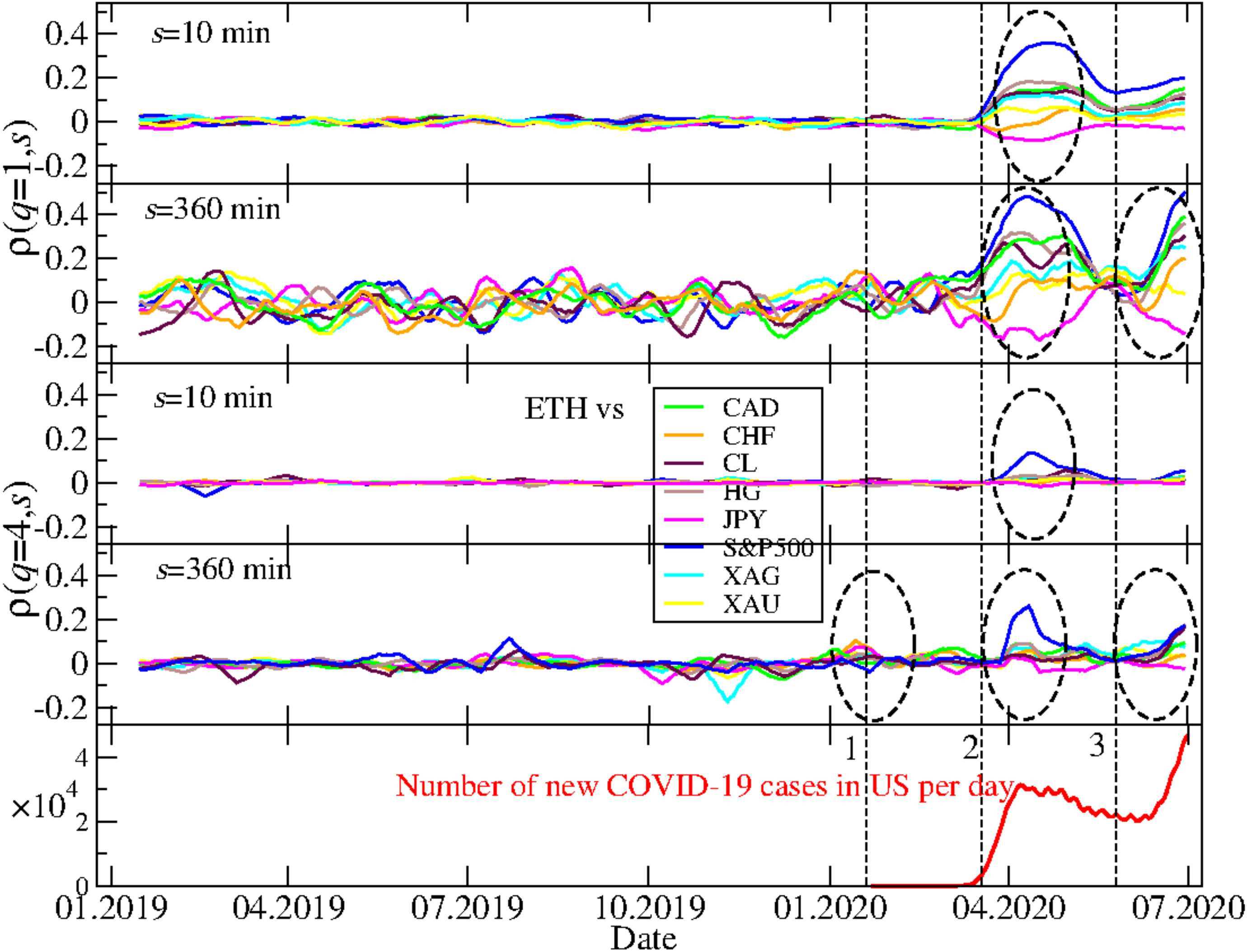}
\caption{Temporal evolution of $\rho(q,s)$ calculated for the ETH/USDT exchange rate and the conventional assets expressed in US dollar: Japanese yen (JPY), Canadian dollar (CAD), Swiss franc (CHF), crude oil (CL), silver (XAG), gold (XAU), copper (HG), and the S\&P500 index. For~more description see caption to Figure~\ref{fig::rhoq.BTC}.}
\label{fig::rhoq.ETH}
\end{figure}

For a comparison, Figure~\ref{fig::rhoq.ETH} shows $\rho(q,s)$ calculated for the ETH/USDT exchange rate and the same conventional assets as in the BTC/USDT case above. We see that the only qualitative difference between Figures~\ref{fig::rhoq.BTC} and~\ref{fig::rhoq.ETH} is a much smaller detrended cross-correlation coefficient value for $q=4$ in the case of ETH/USDT and Event 1 (the first US Covid-19 case). In~fact, the~principal interest was then directed towards BTC and not the other cryptocurrencies. We did not analyze other cryptocurrencies, because only BTC and ETH are traded on the Dukascopy platform, which~all the conventional asset quotes used here came from. If other cryptocurrencies were taken into consideration, they must have been taken from Binance and synchronized additionally, which~might have introduced some spurious correlations. This is why we restricted this analysis to BTC and ETH only.

\subsection{Cryptocurrency Market Structure}
\label{sect::structure}

We have already discussed the fractal autocorrelations of the cryptocurrency exchange rates with respect to US dollar and the cross-correlations between bitcoin and the assets representing conventional markets. Now its time to look at the inner correlation structure of the cryptocurrency market itself. Our data set consists of 128 cryptocurrencies expressed in BTC. This effectively removes the impact of BTC on any other coin, so we have some insight into the market's finer, secondary correlation structure\mbox{ (the primary structure} is such that all the cryptocurrencies are correlated with BTC and form the market as a connected whole~\cite{PhysRep.2020}). In~our earlier work we identified that throughout short history of the market, there were only two cryptocurrencies that played the role of the market's center (in terms of the network centrality): BTC for the most time and ETH in the first half of~2018. ETH,~sometimes together with USDT, was also identified as the most frequent secondary hub of the market, after~BTC~\cite{PhysRep.2020}. Here~we study the market's structure between January~2019 and\mbox{ June 2020---a period }that was not a subject of the previous study.

Minimal spanning tree is an acyclic spanning subset of a complete weighted network that is minimal in terms of the total length of its edges. In~a typical MST construction, the~Pearson correlation coefficient~\cite{Pearson1895} is used to form a correlation matrix that defines a complete network. Here~we follow Refs.~\cite{PhysRep.2020} and~\cite{kwapien2017} and define the network based on the $\rho(q,s)$ matrix. This matrix has entries equal to $\rho(q,s)$ calculated for all possible pairs of the exchange rates X/BTC and Y/BTC, where~X,Y denote any cryptocurrency from our $N=128$ element set. By doing this, we~obtain \mbox{$N(N-1)/2=128*127/2=8128$ }coefficients $\rho(q,s)$ for each choice of $q$ and $s$ (as before, here we restrict our discussion to $q=1$, $q=4$, $s=10$ min, and $s=360$ min). In~order to move to a metric space, we~recalculate the coefficients in a form of a distance:
\begin{equation}
d_{\rm XY}(q,s) = \sqrt{2 \Bigl ( 1 - \rho_{\rm XY}(q,s) \Bigr )}.
\label{distance}
\end{equation}
Since $-1 \le \rho(q,s) \le 1$ for $q>0$, we~obtain limiting values for distance: $0 \le d_{\rm XY}(q,s) \le 2$, where~\mbox{$d_{\rm XY}=0$} means perfect cross-correlation between X and Y, $d_{\rm XY}=2$ means perfect anticorrelation, and $d_{\rm XY}=\sqrt{2}$ means perfect statistical independence. Based on all values of $d_{\rm XY}(q,s)$ we construct MST by using the Prim's algorithm~\cite{prim}.

\begin{figure}[H]%
\centering
\includegraphics[width=0.43\textwidth]{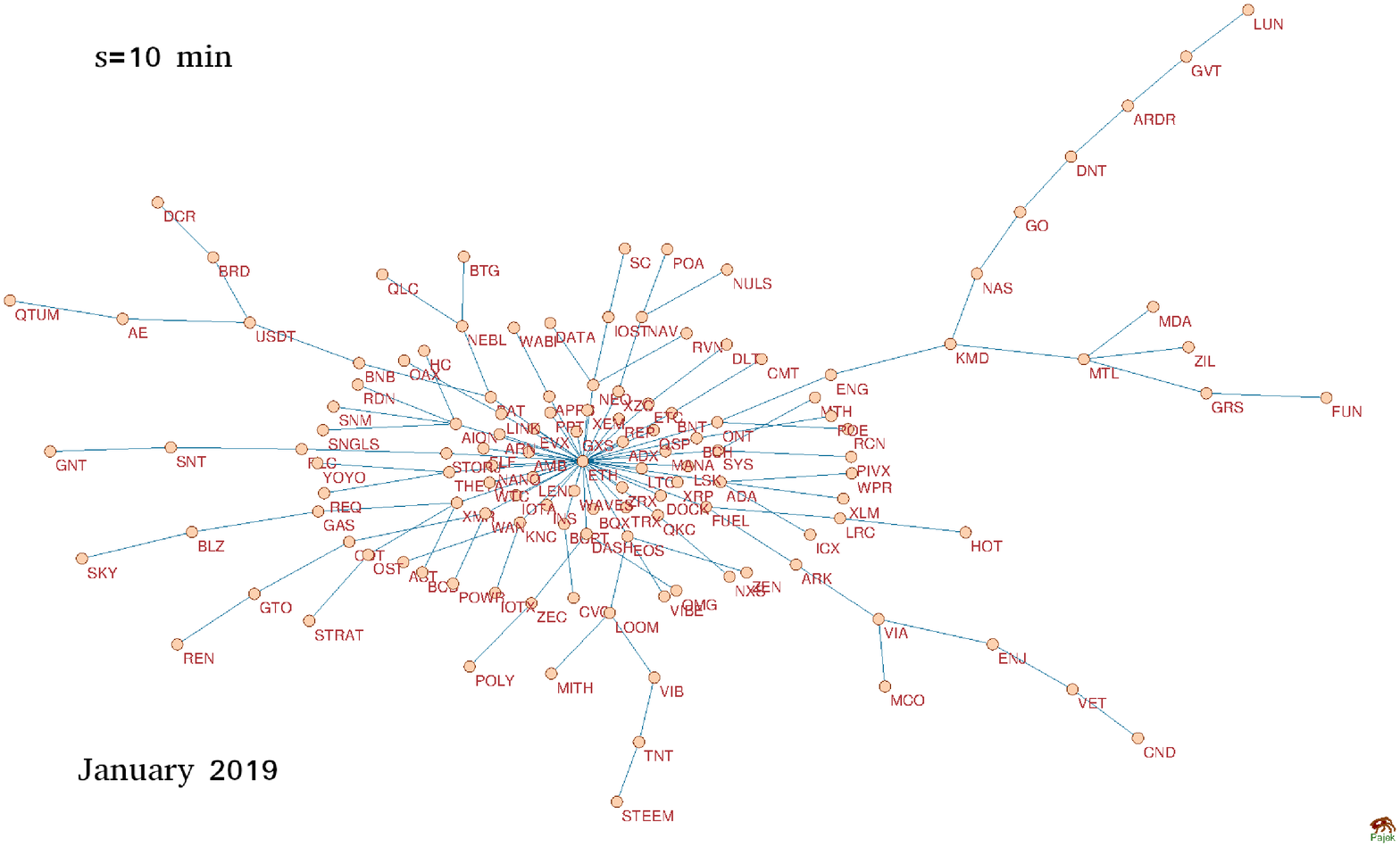}
\includegraphics[width=0.43\textwidth]{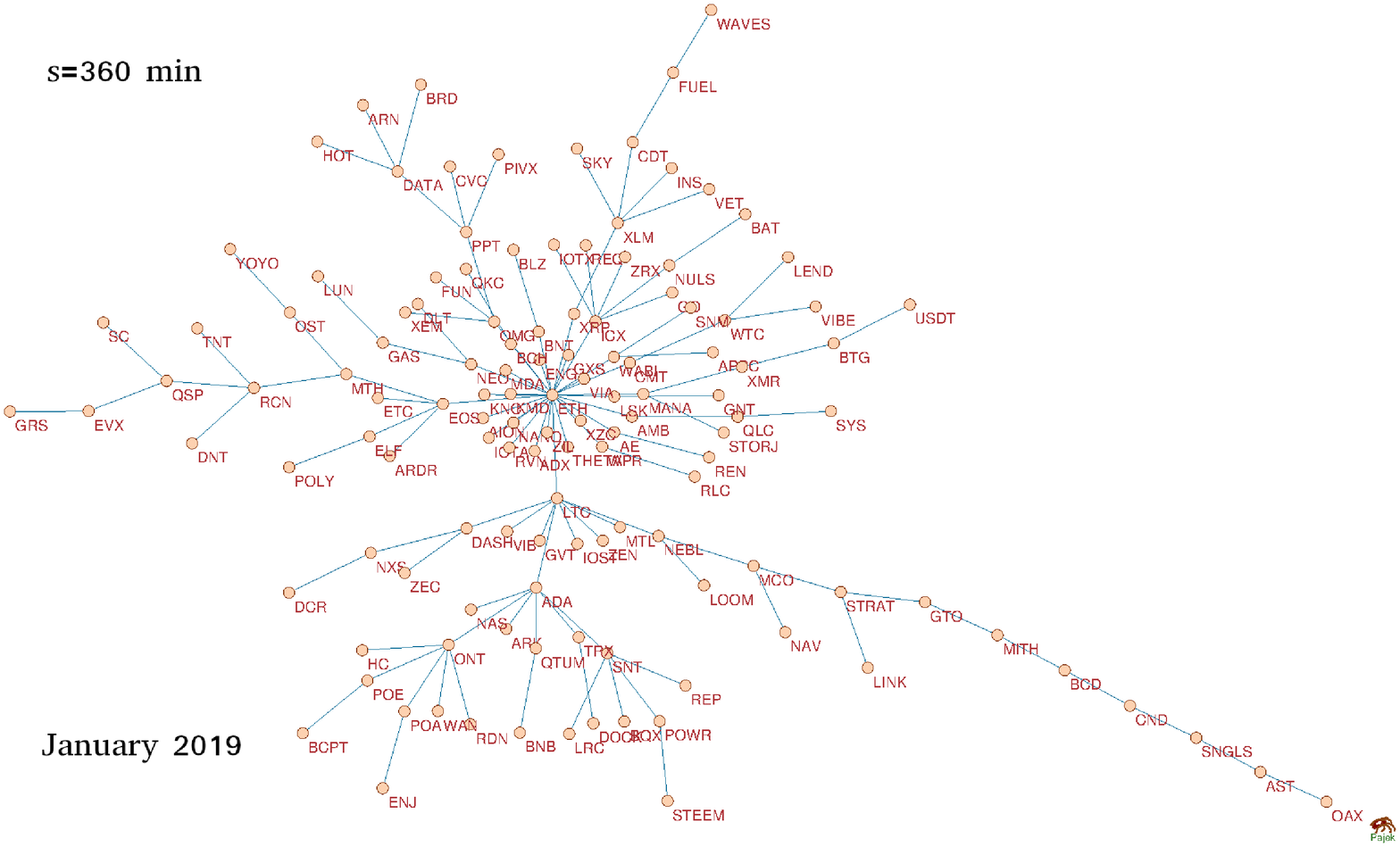}

\includegraphics[width=0.43\textwidth]{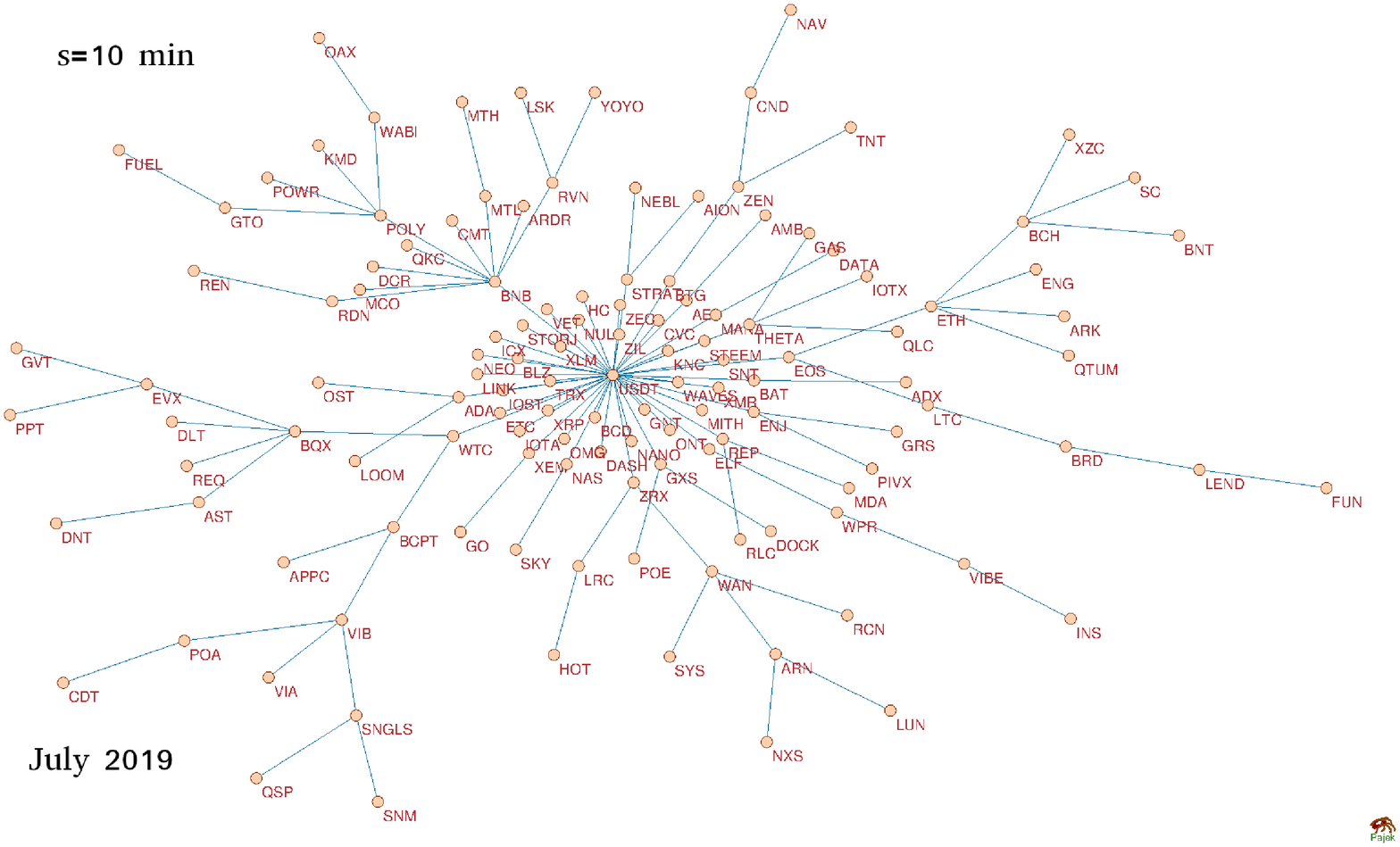}
\includegraphics[width=0.43\textwidth]{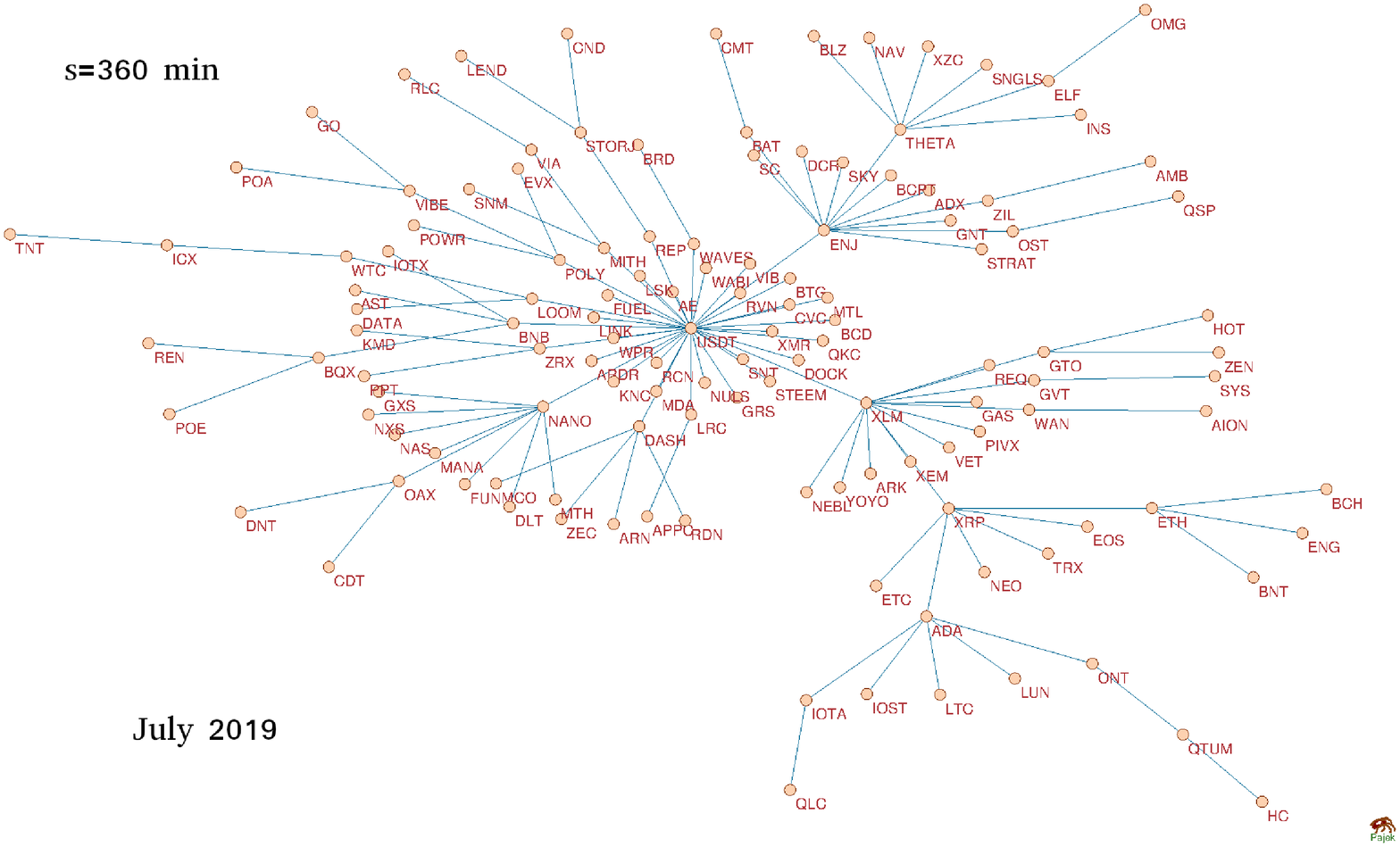}

\includegraphics[width=0.43\textwidth]{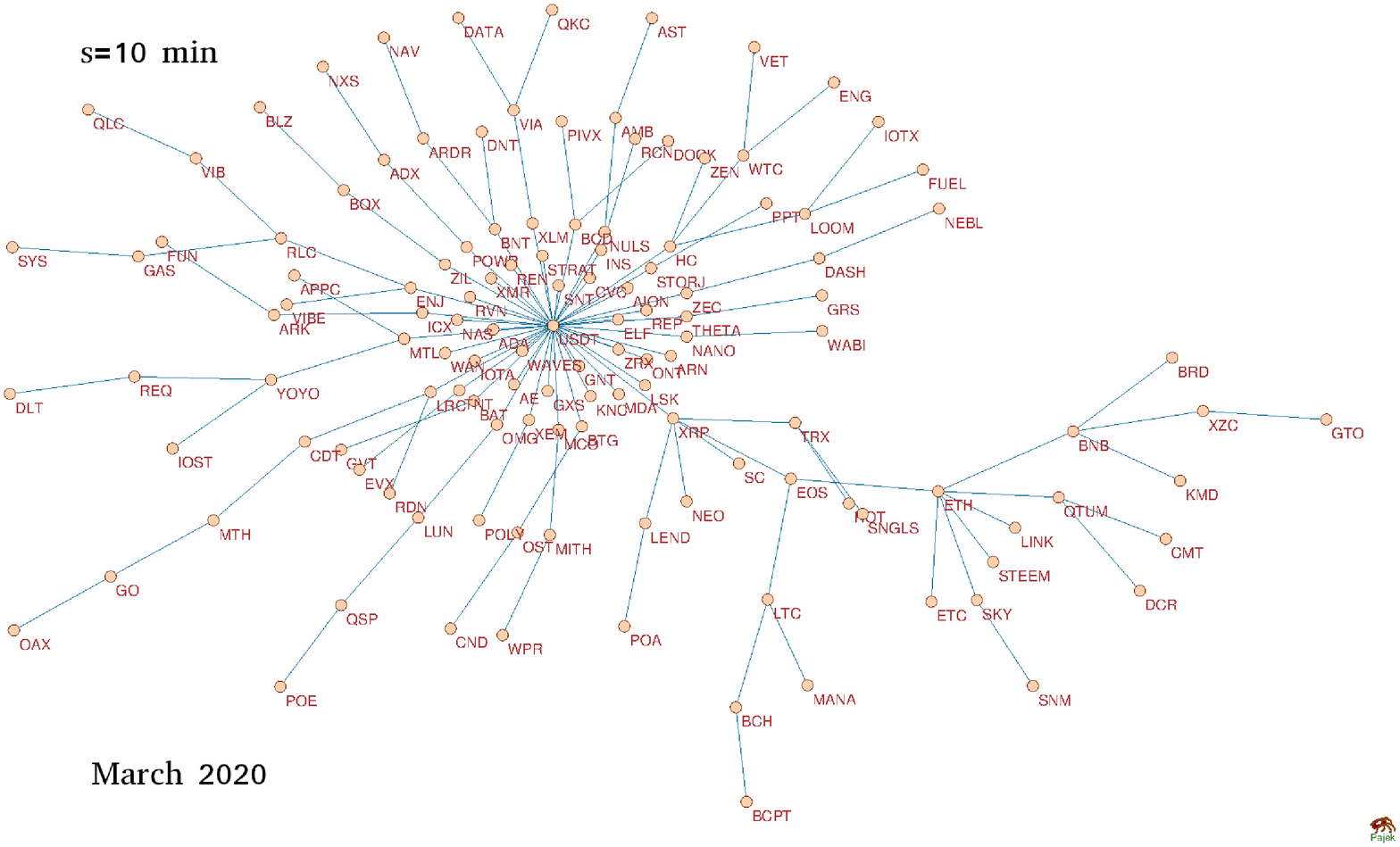}
\includegraphics[width=0.43\textwidth]{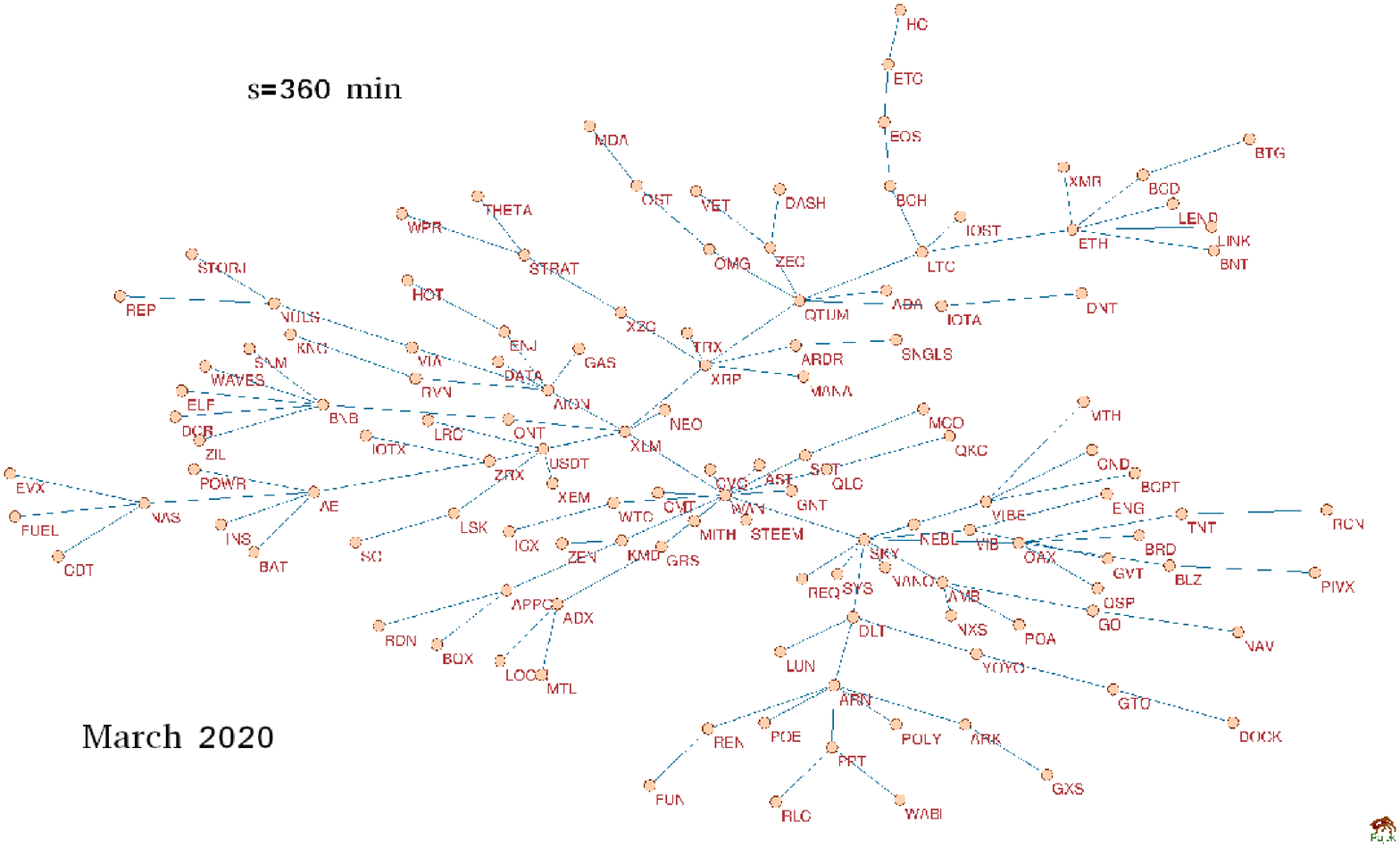}
\caption{Minimal spanning trees (MSTs) calculated based on the $q$-dependent detrended correlation coefficient $\rho(q,s)$ for the exchange rates of a form X/BTC, where~X stands for one of 128 cryptocurrencies traded on Binance~\cite{Binance}. Each node is labeled by the corresponding cryptocurrency ticker. All trees correspond to $q=1$. On the left there are MSTs obtained for $s=10$ min, while~on the right there MSTs obtained for $s=360$ min. Each row shows MSTs calculated in a different period (a 7-day-long moving window with a step of 1 day): January 2019 (top), July 2019 (middle), and March 2020 (bottom).}
\label{fig::MST.q1}
\end{figure}

Figure~\ref{fig::MST.q1} shows $q$MSTs calculated for $q=1$ for three specific periods (from top to bottom): January~2019, July 2019, and March 2020. The~first one was distinguished because it overlaps with a period when ETH was a hub with the highest network centrality (the largest number of connections or the largest degree) for all scales. For~both other periods, a role of the central hub was played by another cryptocurrency: USDT---in July 2019 (all scales) and in March 2020 for short scales (represented by $s=10$ min in Figure~\ref{fig::MST.q1}). However, no overwhelmingly dominant node was observed in MST corresponding to March 2020 and $s=360$ min. In~fact, the~structure of the latter MST differs substantially from the structure of the remaining 5 trees in Figure~\ref{fig::MST.q1}: It can be categorized as a distributed network in contrast to the generally centralized form of the rest, where~there is a clearly identifiable center (ETH or USDT) and the peripheries. There is a possible explanation why USDT becomes a central hub in turbulent periods, especially the sudden dropdowns: Investors that want to close the cryptocurrency positions change them primarily to USDT, which~is a stable coin pegged to USD~\cite{tether} and only then to the proper US dollar. This manoeuvre can mutually correlate most cryptocurrencies via USDT.

\begin{figure}[H]%
\centering
\includegraphics[width=0.48\textwidth]{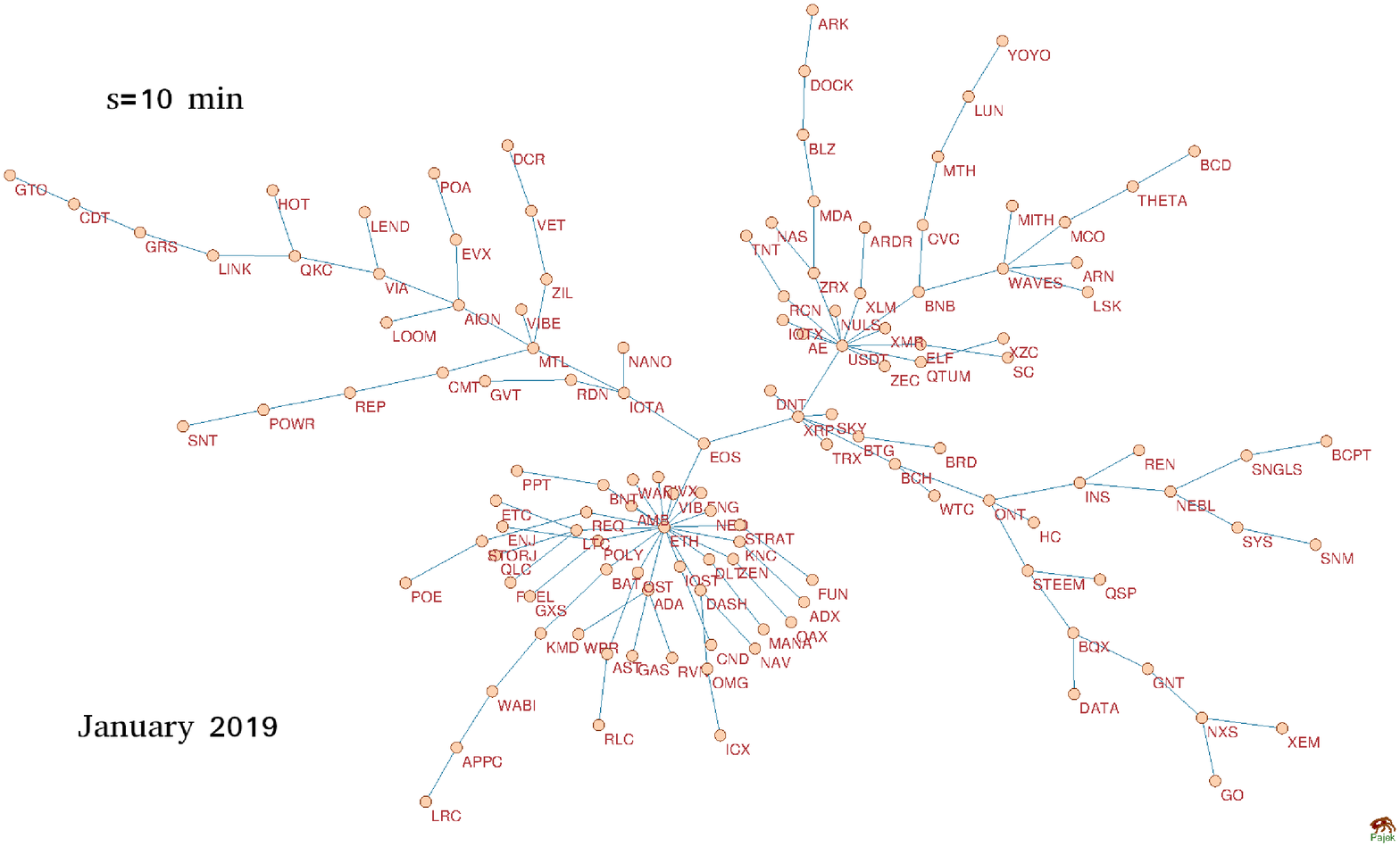}
\includegraphics[width=0.48\textwidth]{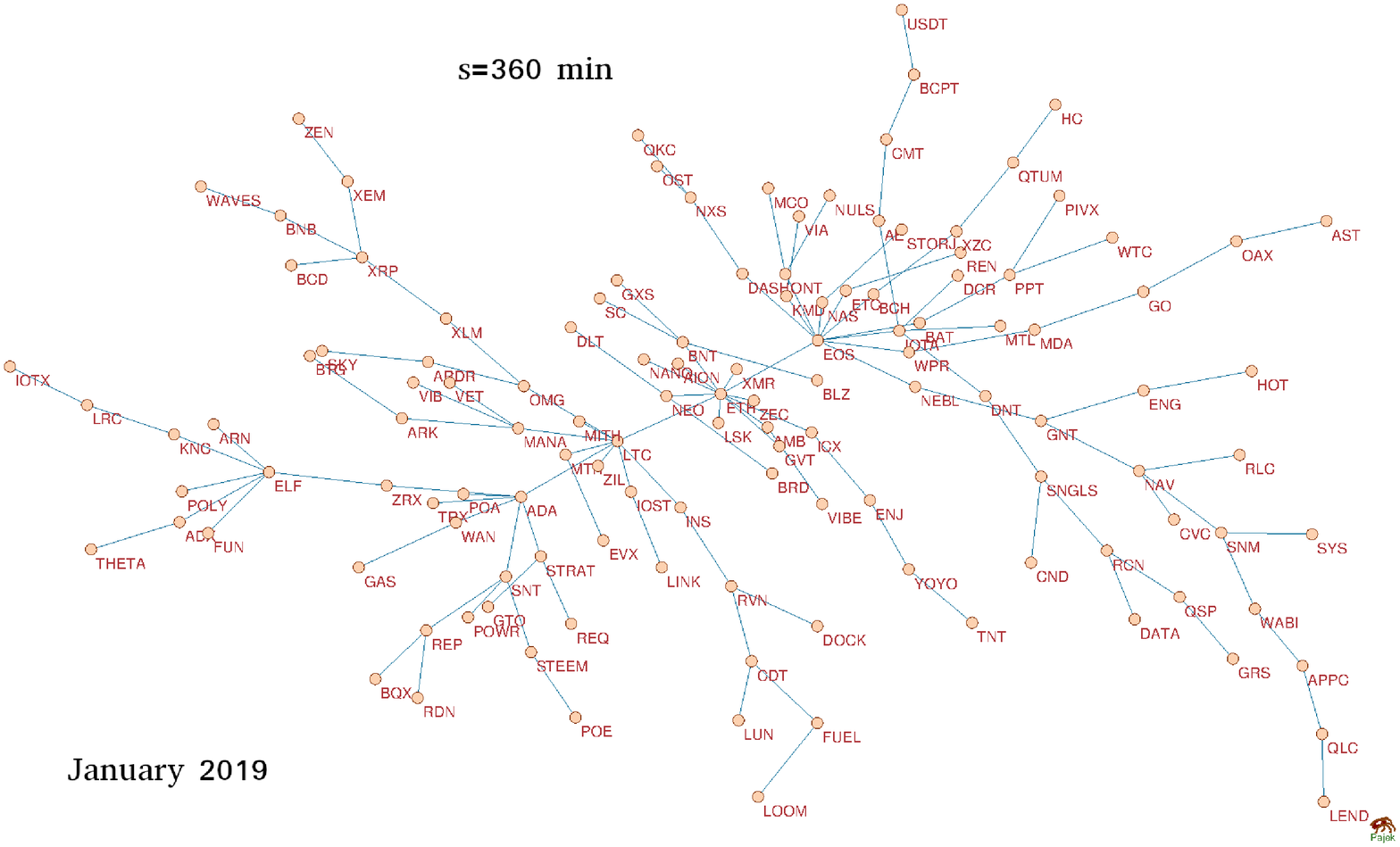}

\includegraphics[width=0.48\textwidth]{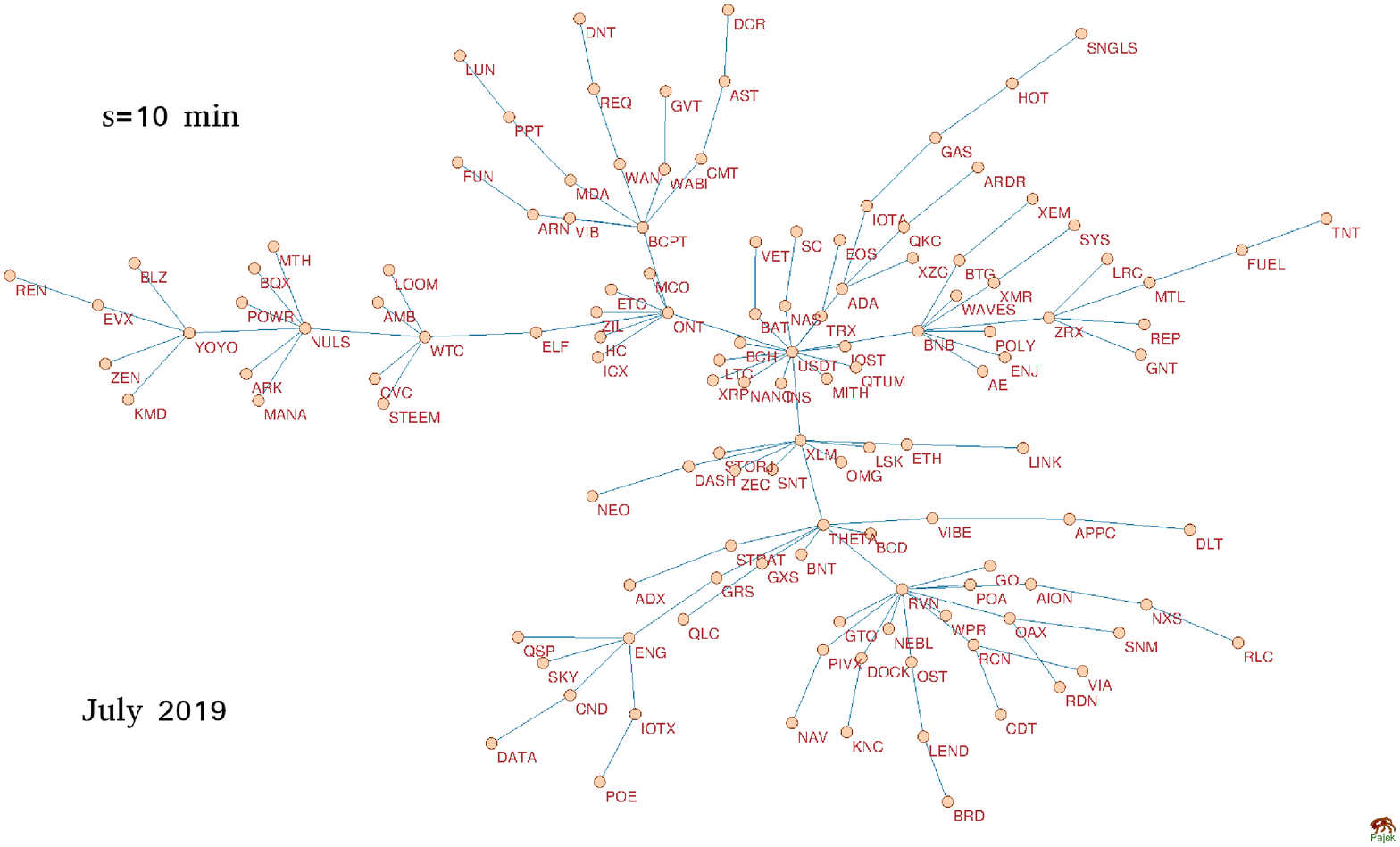}
\includegraphics[width=0.48\textwidth]{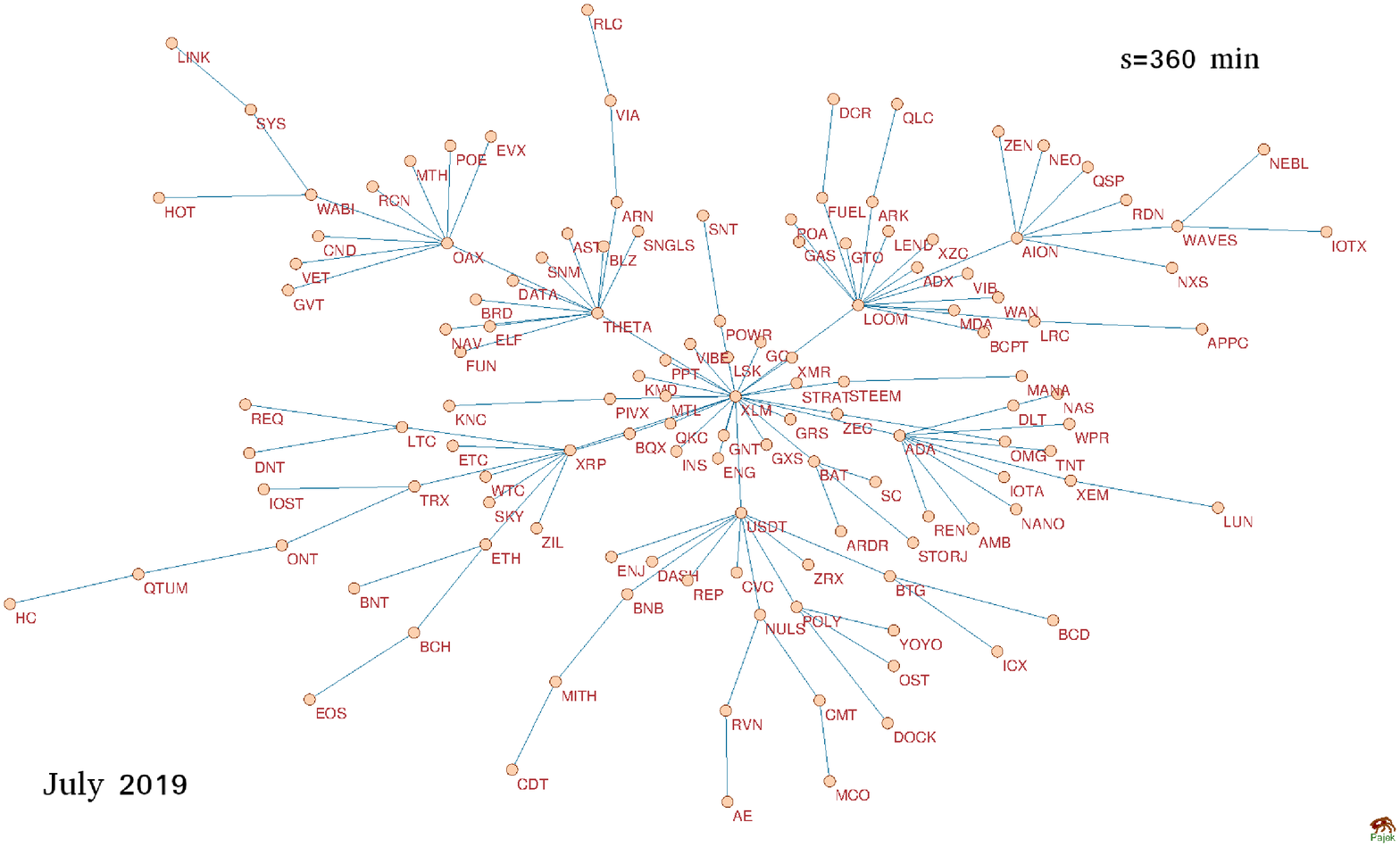}

\includegraphics[width=0.48\textwidth]{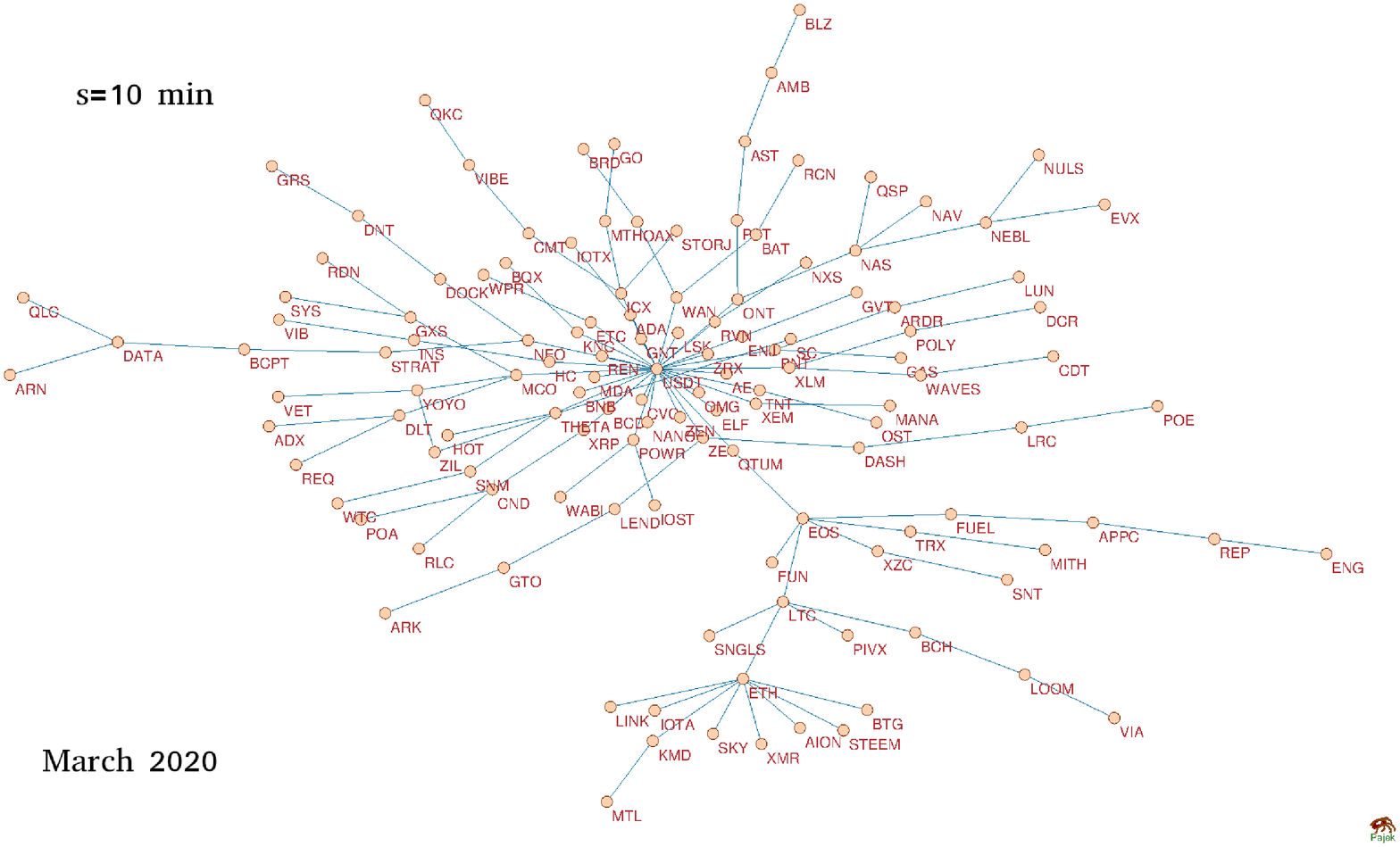}
\includegraphics[width=0.48\textwidth]{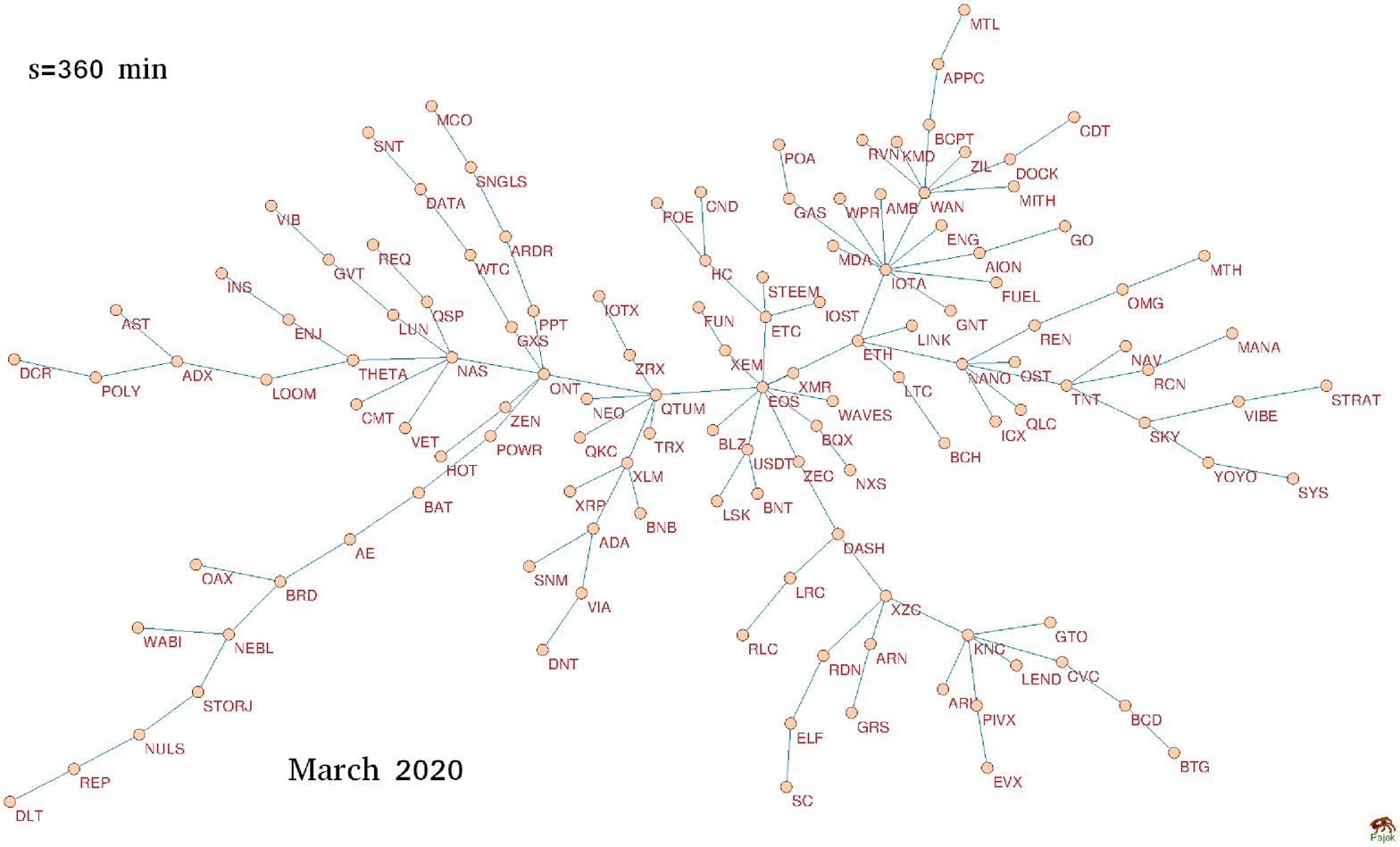}
\caption{Minimal spanning trees (MSTs) calculated based on the $q$-dependent detrended correlation coefficient $\rho(q,s)$ for the exchange rates of a form X/BTC, where~X stands for one of 128 cryptocurrencies traded on Binance~\cite{Binance}. Each node is labeled by the corresponding cryptocurrency ticker. All trees correspond to $q=4$. On the left there are MSTs obtained for $s=10$ min, while~on the right there MSTs obtained for $s=360$ min. Each row shows MSTs calculated in a different period (a 7-day-long moving window with a step of 1 day): January 2019 (top), July 2019 (middle), and March 2020 (bottom).}
\label{fig::MST.q4}
\end{figure}

Now let us consider MSTs constructed from the filtered signals, in which the largest returns were amplified by taking $q=4$ (see Figure~\ref{fig::MST.q4}). In~this case MSTs show a richer pool of forms. Only one tree shows a centralized topology: For $s=10$ min and March 2020, though its central hub (USDT) does not dominate the networks unlike it was for $q=1$ (Figure~\ref{fig::MST.q1}). Moreover, there is only one tree that can be categorized as distributed: For $s=360$ min and March 2020. All the remaining trees reveal intermediate form between the centralized and distributed ones: There are several nodes that can be called local hubs. This is the case of the hierarchical networks that sometimes are scale-free. For~$s=10$~min, such a situation was present in January 2019 (ETH and USDT) and July 2019 (USDT, ONT, XLM, THETA, BCPT, and RVN), while~for $s=360$ min similar situations also occurred in January 2019 \mbox{(ETH, EOS, and LTC)} and in July 2019 (XLM, THETA, LOOM, USDT, ADA, AION, and DAX).

\begin{figure}[H]%
\centering
\includegraphics[width=0.8\textwidth]{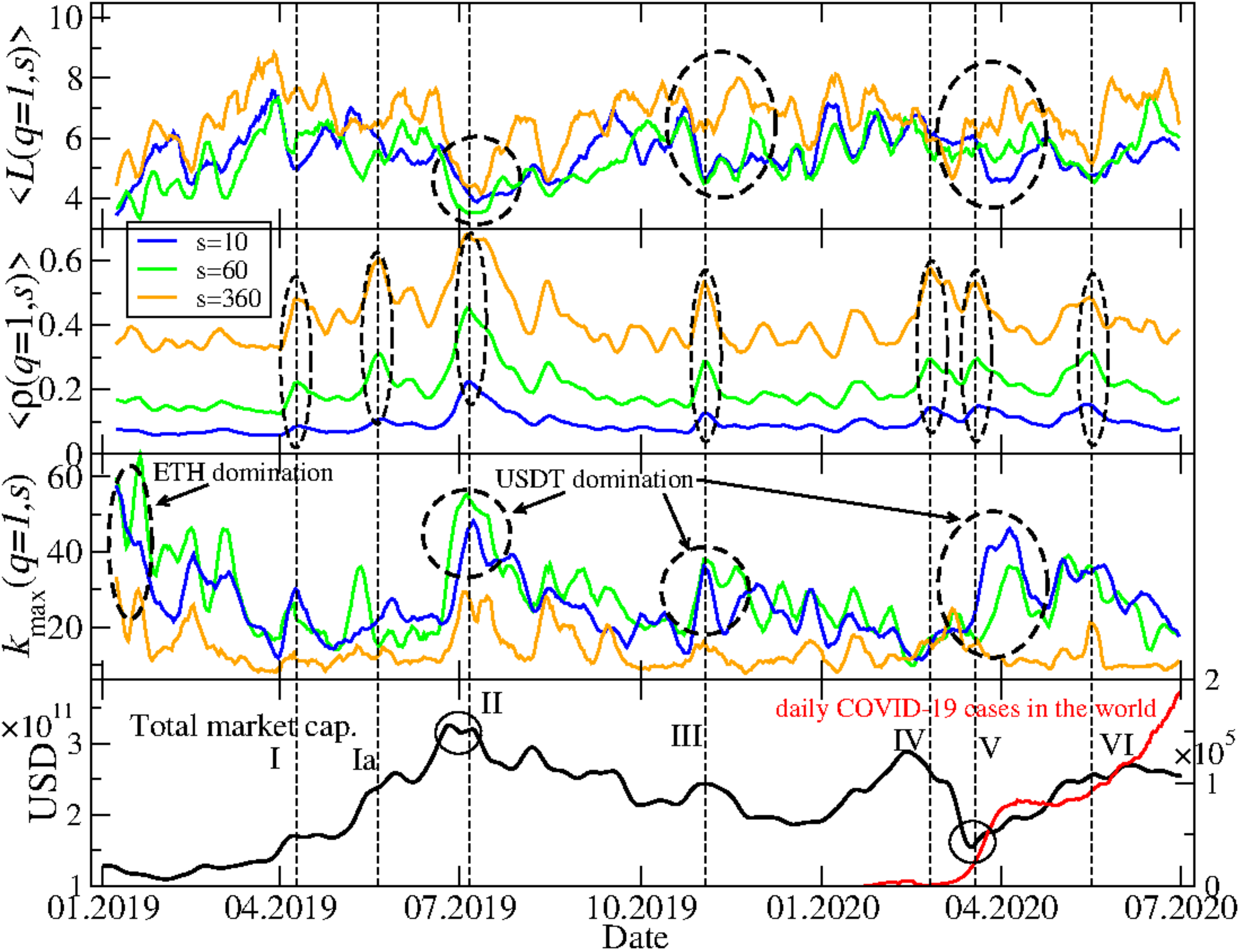}
\caption{Network characteristics describing minimal spanning trees (MSTs) calculated for $q=1$ and for the following scales: $s=10$ min, $s=60$ min, and $s=360$ min. The~average path length $\langle L(q,s) \rangle$ between a pair of MST nodes (top), the~average $q$-dependent detrended cross-correlation coefficient $\langle \rho(q,s) \rangle$ (upper middle), the~maximum node degree $k_{\rm max}(q,s)$ (lower middle), together with the total market capitalization in US dollars and the daily number of new Covid-19 cases in the world (bottom). Several events related to a relatively strong cross-correlations are marked with vertical dashed lines, Roman numerals, and dashed ellipses: Start of a bull market in April 2019 (event I) and its continuation in May 2019 (event Ia), a peak of the bull market in July 2019 (event II), a local peak followed by a sharp drop of the market capitalization in November 2019 (event III), the~Covid-19 panic in mid March 2020 (events IV-V), and the 2nd Covid-19 wave from May 2020 (event VI).}
\label{fig::network.q1}
\end{figure}

\begin{figure}[H]%
\centering
\includegraphics[width=0.8\textwidth]{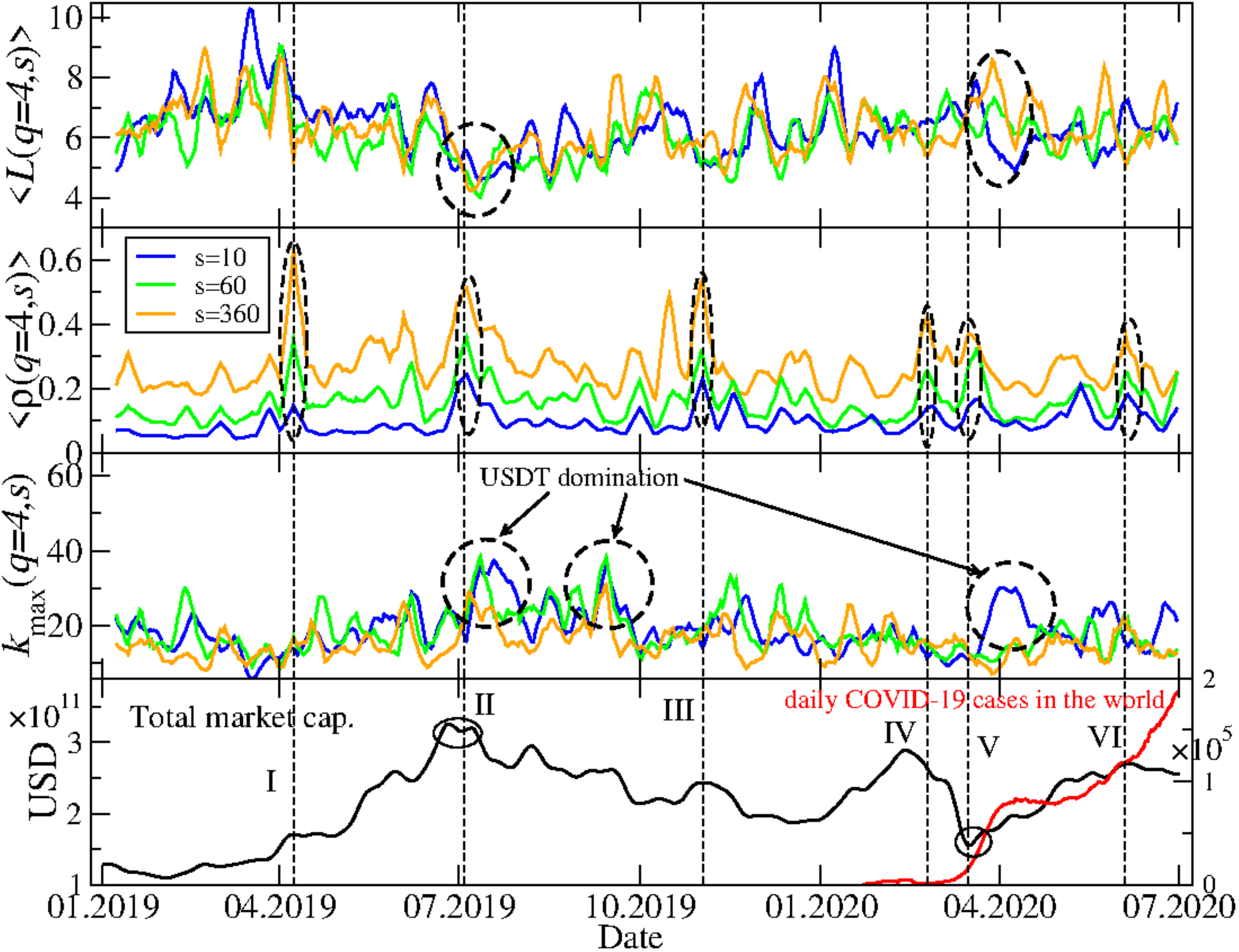}
\caption{The same network characteristics describing MSTs as in Figure~\ref{fig::network.q1}, but~here calculated for $q=4$.}
\label{fig::network.q4}
\end{figure}

The trees shown in Figures~\ref{fig::MST.q1} and~\ref{fig::MST.q4} represent only a few periods, but~in order to look at the market structure evolution over the whole considered interval of time, it is not convenient to look at the trees for individual windows. Therefore, we~calculated a few network characteristics that grasp the essential properties of the MST topology in each window. These are the mean path length $\langle L(q,s) \rangle$ between a pair of the MST nodes (the averaging is carried out over all possible pairs) describing how distributed (large $\langle L(q,s) \rangle$) or concentrated (small $\langle L(q,s) \rangle$) is a tree, the~mean $q$-dependent detrended cross-correlation coefficient $\langle \rho(q,s) \rangle$ (the averaging is carried out over all possible cryptocurrency pairs), describing how strong are typical network edges, and the maximum node degree $k_{\rm max}(q,s)$, describing how central is the main hub. Time evolution of these quantities is shown in Figure~\ref{fig::network.q1} for $q=1$ and in Figure~\ref{fig::network.q4} for $q=4$. Apart from two scales considered in Figures~\ref{fig::MST.q1} and~\ref{fig::MST.q4}, i.e.,~$s=10$ min and $s=360$ min, we~added a medium scale of $s=60$ min.

\unskip

While $\langle \rho(q,s) \rangle$ is largely a different measure than the two other ones, $\langle L(q,s) \rangle$ and $k_{\rm max}(q,s)$ can be related to each other: If $k_{\rm max}(q,s)$ is large, a majority of the nodes is connected to it and $\langle L(q,s) \rangle$ can be small; the opposite relation is also true. Figure~\ref{fig::network.q1} confirms these observations for $q=1$: Typically, the~elevated values of $\langle L(1,s) \rangle$ (top panel) are associated with the suppressed values of $k_{\rm max}(1,s)$ (lower middle panel) no matter what was a particular cause of such a change of the MST structure. The~most important topological changes detectable by $\langle L(1,s) \rangle$ and $k_{\rm max}(1,s)$ occurred after the end of ETH domination in the market in January--February 2019 (topolgy changed from a highly centralized one with ETH being the hub to a rather distributed one), after the end of the bull phase in July--August 2019 (topology returned temporarily to a centralized form but with USDT as the central hub), during a local peak and the subsequent decline of the market in November 2019 (another short period of a centralized topology with USDT domination), and during and after the Covid-19 outbreak March--May 2020 (another phase of USDT domination, but~longer than the preceding ones). On the level of $\langle \rho(1,s) \rangle$, there can be 7 interesting periods pointed out (upper middle panel of Figure~\ref{fig::network.q1}). As one might expect, the~longer scale, the~stronger are the mean cross-correlations; this is a systematical relation throughout the whole analyzed time interval. This is a typical effect observed on many financial markets, which~is related to the liquidity and capitalization differences among the assets. Since the cryptocurrencies with small capitalization are traded less frequently than those with large capitalization, it takes more time for a piece of market information to spread over such cryptocurrencies. Thus, the~cross-correlations among them can only be built and detected on longer scales.

A more interesting situation as regards the different scales $s$ can be found if one compares, on the one hand, $k_{\rm max}(1,s)$ between these scales and, on the other hand, $\langle L(1,s) \rangle$. Let us look at two events: A peak and decline of the bull market in July 2019 (event II) and the Covid-19 pandemic (events IV--VI). During the former, $k_{\rm max}(1,s)$ shows a standard behaviour, i.e.,~for $s=10$ min and $s=60$~min it is significantly larger than for $s=360$ min; the same can be said for the events IV-VI. According with what it has been said above, we~might expect that in both cases $\langle L(1,360~{\rm min}) \rangle$ should be larger than $\langle L(1,10~{\rm min}) \rangle$ and $\langle L(1,60~{\rm min}) \rangle$. while~this was the case, indeed, during the pandemic outbreak in March 2020, nothing like this happened during the bull market peak in July 2019, when $\langle L(1,s) \rangle$ was comparable for all the scales. Such a deviation from the overall rule that a longer scale is associated with a better-developed hierarchical or a more distributed MST topology (smaller $k_{\rm max}(q,s)$) and a shorter scale is associated with either a more centralized network topology (larger $k_{\rm max}(q,s)$) was rather unusual as for the whole studied period.

Figure~\ref{fig::network.q4} differs from Figure~\ref{fig::network.q1} only in that it shows the same quantities but for $q=4$ (mainly the cross-correlations for the large-amplitude returns are considered now). The~above-discussed relation between $k_{\rm max}(q,s)$ for different scales $s$ is less clear for $q=4$ than it is for $q=1$. Only in July 2019 and in March-April 2020 there can be distinguished some characteristic structures in time evolution of $k_{\rm max}(4,s)$ and $\langle L(4,s) \rangle$. For~the events that took place in July 2019, a relation between values of the maximum node degree for different scales and a relation between values of the mean path length also for different scales resemble those identified for $q=1$. For~the Covid-19 outbreak period, the~small difference is that now $k_{\rm max}(4,60~{\rm min})$ is comparable to its counterpart for $s=360$ min instead of $s=10$ min as for $q=1$. There is no difference between $q=1$ and $q=4$ as regards  $\langle \rho(4,s) \rangle$: The longer the scale is, the~stronger are the cross-correlations.

To summarize observations related to the MST topology, in the analyzed period from January 2019 to June 2020 this topology used to change substantially during periods of large volatility in such a way that from a hierarchical or distributed network structure that was typical outside these periods it used to transform itself to a more centralized structure with a dominating hub and much stronger cross-correlations between the nodes (see also~\cite{medina2020}). The~most interesting period was the Covid-19 pandemic, during which on short and moderate scales for $q=1$ one observed first a significant increase of the MST centralization (large $k_{\rm max}(q,s)$) and a subsequent slow return to a more distributed form (moderate $k_{\rm max}(q,s)$) but still with a distinguished central hub. However, on the longest scale this effect was not observed and $k_{\rm max}(1,360~{\rm min})$ was elevated only once in May 2020. This suggests that the most sudden and nervous movements that correlate the market and centralize its topology on short time scales tend to be blurred as time passes and we go from short to long scales, where~topology becomes much more of a distributed or hierarchical type. Such a behaviour observed recently during the pandemic, which~can be considered as an external perturbation to the market, differs from the behaviour observed during the peak and collapse of the bull market in July 2019, which~was no doubt a result of the internal evolution of the market. Whether this internal/external events may be source of the observed peculiarities of the Covid-19 period, one cannot state for sure as both events were unique during the analyzed time interval and cannot be confirmed by other events of similar type.


\section{Summary}

In our work we focused on dynamical and structural properties of the cryptocurrency market. We analyzed empirical data representing the exchange rates of 129 cryptocurrencies traded on the Binance platform, including BTC. The~analysis comprised three parts, each of which was intended for investigating a different aspect of the market structure. We started from a multifractal analysis of the BTC/USDT exchange rate as the most important one together with a similar analysis of an artificial cryptocurrency index based on 8 the most capitalized coins. This analysis may be considered as an extension of the analysis reported in Ref.~\cite{PhysRep.2020} on the most recent time interval from January 2019 to June 2020. The~results showed that throughout this interval the cryptocurrency dynamics produces multifractal fluctuations (returns) with some intermittent signatures of bifractality that can be assigned to specific volatile periods like the Covid-19 outburst in March 2020 or a bull market start in April 2019. Moreover, on a level of the return distributions such bifractal-like singularity spectra can be accounted for by the pdf/cdf power-law tails that fall into the L\'evy-stable regime~\cite{EPL2010}. Outside these volatile periods spectra are wide but with much smaller left-right asymmetry.

The analysis of the cross-correlations between the cryptocurrency market represented by BTC/USD or ETH/USD and the conventional markets represented by the major fiat currencies, the~most important commodities (e.g., crude oil and gold), and the US stock market indices brought us to an observation that the cryptocurrency market was decoupled from the remaining markets throughout the whole year 2019, but~it used to couple temporarily to those markets during some events in the first half of 2020, like in January when the first Covid-19 case was reported in the United States, in March during the pandemic outbreak, and in May-July during the pandemic's 2nd wave. In~the first case, BTC was anticorrelated with the major stock market indices like S\&P500 and Nasdaq100, but~in the second and the third cases the analogous cross-correlations were positive. Positive were then also the cross-correlations between BTC and several fiat currencies and commodities. A lack of the statistically valid cross-correlations in 2019, when the conventional assets did not experience anything turbulent, was supposedly caused by the asymmetry in market capitalization between the cryptocurrency market and the conventional markets to the disadvantage of the cryptocurrency market, which~was too small to have any sizeable impact on the other markets. However, the~conventional markets can easily influence the cryptocurrency market if they are turbulent. This is exactly what was observed in March 2020 and June 2020. Except for January 2020, when, unlike BTC/USD, ETH/USD was not correlated with the conventional assets, both the exchange rates reveal a similar relation with these assets.

A network representation of the cryptocurrency market can shed light on the market's inner cross-correlation structure. Our analysis based on the exchange rates of 128 coins with respect to BTC revealed that turbulent periods on the market result in a sudden transition between different network topology types. During the periods of normal dynamics, the~market has a distributed-network topology or a hierarchical-network topology, in which no node dominates the network and there is a hierarchy of hubs with decreasing centrality (e.g., node degree). Typically, for long scales the hierarchical-network topology is more pronounced than for short scales, where~a centralized-network topology prevails. This is because the cryptocurrencies of small capitalization are less liquid, so a piece of information needs more time to be fully processed by them and the cross-correlations, especially those more subtle, sector-like, and related to less prominent cryptocurrencies, can only build up on sufficiently long time scales. This picture is altered if there comes a volatile period. During such periods the network becomes highly centralized with one dominating hub for all the scales. The~most often the role of such a hub is played by USDT, because it is pegged to the US dollar and, thus, considered as more stable than other cryptocurrencies. If investors flee the cryptocurrency market, they first change their assets to USDT, and only then to USD, which~can correlate a majority of the cryptocurrencies together via USDT. However, a compact, star-like topological form exists shortly and soon it returns to a more distributed, more branched form.

We also noticed that the most significant events as regards their impact on the market topology---the transition from a bull market to a bear market in July 2019 and the Covid-19 pandemics that started in March 2020---differ in some details of that impact. During the pandemics, a transition from a centralized form to a distributed form occurred predominantly on short and medium scales, while~on long scales it was less pronounced. Contrary to that, in July 2019 the topology shift was visible on all the scales. It is a matter of future analyses to address a question whether this difference can be related to endogenous (a trend reversal) vs. exogenous (the pandemic) origin of both events.

\vspace{6pt} 

\authorcontributions{Conceptualization, S.D., P.O., and M.W.; methodology, S.D., J.K., P.O., T.S., and M.W.; software, T.S. and M.W.; validation, S.D., J.K., P.O., T.S., and M.W.; formal analysis, P.O., T.S., and M.W.; investigation, P.O., T.S., and M.W.; resources, T.S. and M.W.; data curation, T.S. and M.W.; writing--original draft preparation, J.K. and M.W.; writing--review and editing, J.K. and M.W.; visualization, T.S. and M.W.; supervision, S.D. and P.O. All authors have read and agreed to the published version of the manuscript.}

\conflictsofinterest{The authors declare no conflict of interest.} 

\reftitle{References}

\end{document}